\documentclass{elsart}

\usepackage[dvips]{graphics}

\begin{document}

\begin{frontmatter}
\title{CALIBRATION OF SUPER-KAMIOKANDE USING AN ELECTRON LINAC}

\collab{The Super-Kamiokande Collaboration}

\author[icrr]{M. Nakahata\thanksref{corresponding}}, 
\author[icrr]{Y. Fukuda}, 
\author[icrr]{T. Hayakawa}, 
\author[icrr]{E. Ichihara}, 
\author[icrr]{K. Inoue},
\author[icrr]{K. Ishihara}, 
\author[icrr]{H. Ishino}, 
\author[icrr]{Y. Itow},
\author[icrr]{T. Kajita}, 
\author[icrr]{J. Kameda}, 
% ...

\address[icrr]{Institute for Cosmic Ray Research, University of Tokyo, Tanashi,
Tokyo 188-8502, Japan}
% ...

\thanks[corresponding]{Corresponding author: Kamioka Observatory, 
Higashi-Mozumi, Kamioka-cho, Yoshiki-gun, Gifu-ken, 506-1205 Japan, 
Tel: +81 578 5 9603, Fax: +81 578 5 2121, e-mail nakahata@icrr.u-tokyo.ac.jp} 
\thanks[martens]{Present address: Department of Physics and Astronomy, 
SUNY at Stony Brook, NY 11794-3800, USA}
\thanks[halverson]{Present address: NASA, JPL, Pasadena, CA 91109, USA}
\thanks[flanagan]{Present address: Accelerator Laboratory, 
High Energy Accelerator Research Organization (KEK), Japan}
\thanks[conner]{Present address: Enrico Fermi Institute,
University of Chicago, Chicago, IL 60637, USA}
\thanks[hayato]{Present address: 
Institute of Particle and Nuclear Studies,
High Energy Accelerator Research Organization (KEK), Japan}
\thanks[poland]{Supported by the Polish Committee for Scientific Research }
\thanks[wai]{Present address: Department of Physics, Stanford University, 
CA 94305, USA}

\newpage
\begin{abstract}
In order to calibrate the Super-Kamiokande experiment for solar neutrino
measurements, a linear accelerator (LINAC) for electrons was installed at 
the detector.
LINAC data were taken at various positions in the detector volume, 
tracking the detector response in the variables relevant to solar neutrino 
analysis. 
In particular, the absolute energy scale is now known 
with less than 1\% uncertainty. 
\end{abstract}


\end{frontmatter}

\newpage
\section{Introduction}

Past solar neutrino 
experiments\cite{ref:cl,ref:kam,ref:sag,ref:gal}
have established the solar neutrino problem. 
The data collected by these experiments suggest an energy dependent suppression 
of the solar neutrino flux measured on Earth. 
Some physical interpretations of the data make specific predictions for 
spectral distortions of the measured solar neutrino spectrum\cite{ref:hl}. 
Thus a new generation of solar neutrino experiments set out to perform high 
precision measurements of the solar neutrino spectrum. 
On April 1, 1996, Super-Kamiokande (SK) was the first of these experiments to 
start taking data\cite{ref:sksol}. 

SK uses the elastic scattering of electrons in water to observe 
the high energy $^8$B neutrinos from the sun. 
As a water Cherenkov counter it provides real time directional and energy 
information on the recoil electron from the neutrino interaction. 
Since the energy of solar neutrinos is less than $20\,\rm{MeV}$ and angular 
resolution is limited by multiple Coulomb scattering, a kinematic 
reconstruction of the incident neutrino's energy is precluded. 
The recoil electron energy measured in the detector only gives a lower 
limit for the energy of the incident neutrino, so the features of the incident 
solar neutrino spectrum have to be inferred from the resulting 
recoil electron spectrum measured in the detector. 
This increases the sensitivity of spectral analysis and $^8$B solar neutrino 
flux estimation to the calibration of energy scale and resolution in the 
detector. 
A linear accelerator (LINAC) was installed at SK to provide 
precise detector calibration with single electrons of known energy and direction 
at various positions in the detector volume. 

\section{Super-Kamiokande and Solar Neutrinos}  \label{secSK}

SK is a water Cherenkov detector located in 
the Kamioka Mine in Japan. 
It is divided into an inner and an outer detector (ID and OD), 
which are concentric cylindrical water volumes separated by an optical barrier. 
All photomultiplier tubes (PMTs) collecting Cherenkov light are mounted on 
this optical barrier, with 20~inch ID PMTs looking inward and 8~inch OD PMTs 
looking outward. 
The OD has a uniform thickness of 2.5~m surrounding the ID, and in the solar 
neutrino analysis it is used as a veto counter. 
It also passively shields the ID from gamma activity from the surrounding rock. 
Cherenkov light emitted by electrons recoiling from neutrino scattering in 
the ID is collected by 11,146 ID PMTs. 
These PMTs are uniformly distributed on a 0.707~m square grid, enclosing 
32,000~metric~tons of water in a volume of 36.2~m height and 33.8~m diameter. 
The fiducial volume for solar neutrino analysis starts 2~m inside the 
physical confines of the ID, and contains 22,500~metric~tons of water. 
The timing range of the time-to-digital converters 
used with the ID PMTs is 1.2~$\mu$s, while 
the time needed for light to travel the diagonal of the ID is about 230~ns. 

Solar neutrinos measured in SK have energies from 5 to 18~MeV. 
Their recoil electrons have ranges of a few centimeters, allowing use of a 
point fitter for vertex reconstruction. 
Knowing the vertex position, the characteristic directionality of Cherenkov 
radiation allows reconstruction of the recoil electron direction from the 
distribution of light around this vertex. 
The total amount of light emitted in an event is used in the energy 
determination. 
With a yield of about 6 hit PMTs per MeV of electron energy, even for the 
highest energy solar neutrino events less than 1\% of all PMTs have signals. 
The number of hit PMTs is related to the electron energy. 
This number must be corrected for absorption and scattering as well as for 
geometrical acceptance, depending on event location and direction in the 
detector.

SK's predecessor, 
Kamiokande, used gamma-rays from the Ni(n,$\gamma$)Ni reaction for the 
calibration of its absolute energy scale\cite{ref:kful}.
Uncertainties in branching ratios and the neutron absorption cross sections for 
different nickel isotopes limit the accuracy of such an energy calibration to 
1--2\%. 
Nickel calibration also provides no information on the angular resolution 
of the detector and only limited information on energy resolution. 
The angular resolution as a function of energy is used to fit the distribution 
of event directions with respect to the direction to the Sun. 
This fit is fundamental to the solar neutrino analysis in SK. 

The LINAC offers the means to study the detector response to electrons, its 
position dependence and angular resolution {\it in situ}. 
It allows the injection of single electrons of well-controlled energy at various 
positions in the ID. 
The LINAC covers the energy range of solar neutrino events and provides an 
excellent calibration of the absolute energy scale. 

The following sections will describe the LINAC and its beam transport system, 
its calibration, and the results obtained from the first scan of the ID with 
this new and powerful calibration tool. 

\section{The Electron LINAC}  \label{secEL}

\begin{figure}
\begin{center}
\resizebox{10cm}{!}{\includegraphics{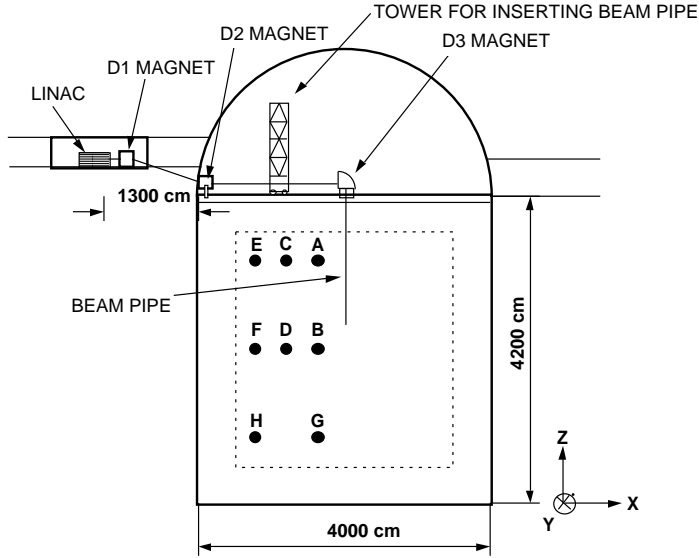}}
\end{center}
\caption{The LINAC and its beam line at the SK detector. 
The fiducial volume for the solar neutrino measurement is indicated by a
dashed line.
Black dots indicate where in the fiducial volume calibration data were taken 
with the LINAC (see also Tab.~\protect{\ref{t:dpos}}).}
\label{f:gbl}
\end{figure}

The LINAC employed at SK is a Mitsubishi ML-15MIII produced
for medical purposes in 1978. 
It was used at the hospital of Miyazaki Medical University until 
its installation at the SK detector in 1996. 
Certain modifications were necessary to adapt the LINAC to its new purpose. 
It is now mounted on a solid support structure in a tunnel to the side and 
slightly above the top of the steel tank containing the 
water Cherenkov detector (Fig.~\ref{f:gbl}). 
For SK calibration, single electrons are needed in the detector. 
A special electron gun reduces the number of electrons 
entering the acceleration tube to appropriate levels. 
Its output current is adjustable, allowing control of the beam intensity. 

Microwave pulses of $\sim2\,\mu$s width are generated in a Mitsubishi PV2012M 
klystron with an adjustable pulse rate between 10 and 66~Hz. 
Electrons from the electron gun 
are accelerated as they travel with the microwave in the accelerating tube. 
Manipulating the input power and frequency of the microwave changes the 
average beam momentum. 
The electron energy can be varied in a range from 5 to 16~MeV, well 
matched to recoil electron energies from solar neutrinos. 
Features of the LINAC are summarized in Tab.~\ref{t:linac}.

\begin{table}
\begin{tabular}{|l|l|}  \hline
Type & Mitsubishi ML-15MIII \\
Accelerating tube & 1.69~m length 26~mm inner diameter \\
Acceleration type & traveling-wave \\
Microwave frequency & 2.856 GHz \\
Klystron & Mitsubishi PV2012M \\
Electron gun & 0.125~mm diameter tungsten filament \\
Electron gun intensity & 200 $\mu$A maximum \\
Vacuum in accelerating tube & 10$^{-7}$ torr \\
Maximum beam intensity & $\sim$10$^{6}$ electrons/pulse at the
accelerator tube end\\
Beam momentum & 5 - 16 MeV/c \\
Pulse width & 1 -- 2 microsecond \\
Repetion rate & 10 -- 66 pulses / second \\
Beam size & $\sim$ 6 mm \\
Beam angular spread & $\sim$ 3 mrad \\
Power consumption & 30 KVA \\
\hline
\end{tabular}
\caption{Technical data on the electron LINAC.}
\label{t:linac}
\end{table}

\begin{figure}
\begin{center}
\resizebox{10cm}{!}{\includegraphics{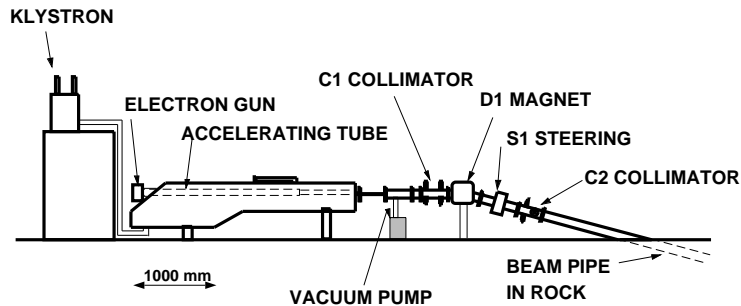}}
\end{center}
\caption{Beamline detail: First bending magnet (D1) and associated collimators. 
Here the beam momentum is defined.}
\label{f:trnsd1}
\end{figure}

\begin{figure}
\begin{center}
\resizebox{10cm}{!}{\includegraphics{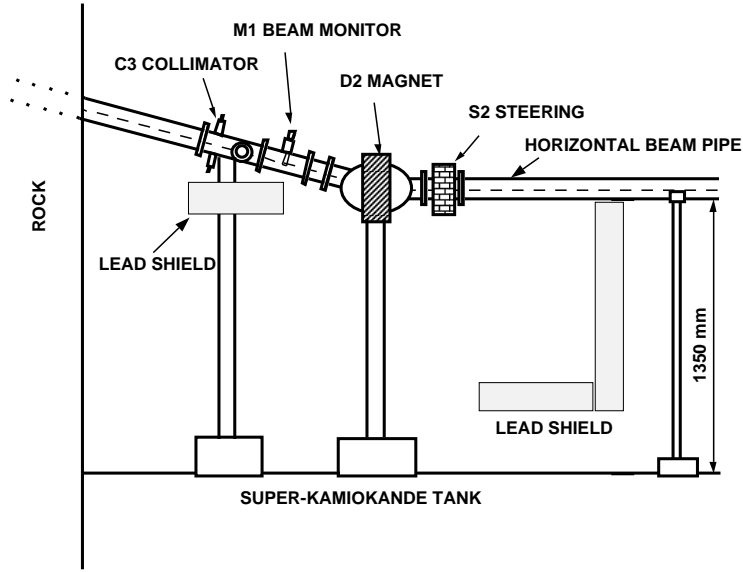}}
\end{center}
\caption{Beamline detail: Returning the beam to horizontal after momentum 
selection.}
\label{f:trnsd2}
\end{figure}

After the accelerating tube, the electron beam is rather divergent and 
spans a modest momentum range. 
Mono-energetic electrons are selected from this spectrum by an arrangement 
of collimators surrounding D1, the first 15~degree bending magnet 
(Fig.~\ref{f:trnsd1}). 
After the C3 collimator (Fig.~\ref{f:trnsd2}), the beam momentum spread is 
reduced to 0.5\% at FWHM. 
Constraining beam momentum and divergence reduces the beam intensity from 
$\sim10^6$ to a few electrons per microwave pulse. 
Thus almost the entire beam intensity is either dumped into 
collimators or deflected out of the beamline by the magnet. 
If any gammas generated in this process were to reach the ID, additional light 
from their Compton electrons would produce correlated background, altering the 
energy calibration. 

To shield the detector from this radiation, the beam pipe passes through 9~m of 
rock after the 15~degree downward bend in D1 before it emerges on top 
of the SK detector. 
There it is bent back to horizontal by D2, also a 15~degree bending magnet 
(Fig.~\ref{f:trnsd2}). 
Gammas travelling towards SK in the inclined section of the beampipe before 
D2 are absorbed in a lead shield. 

\begin{figure}
\begin{center}
\resizebox{14cm}{!}{\includegraphics{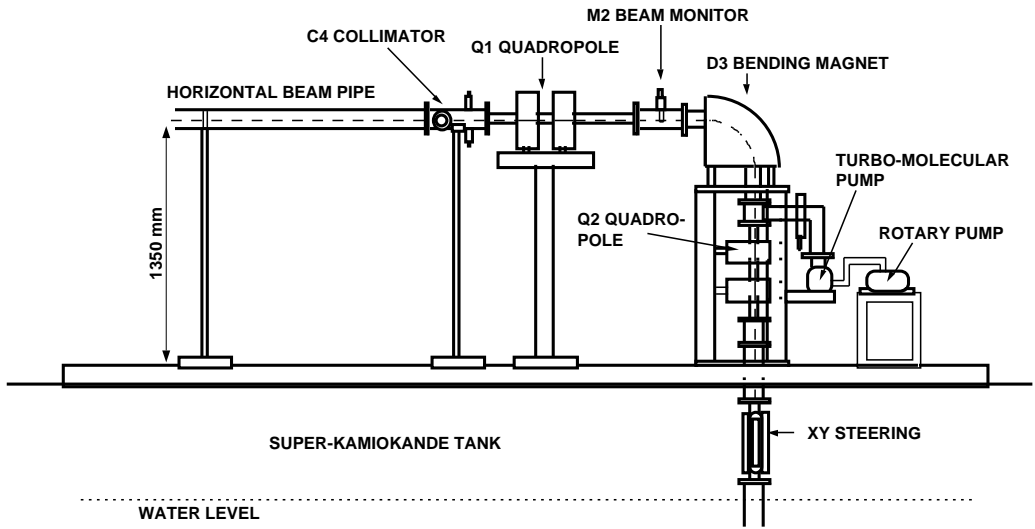}}
\end{center}
\caption{Beamline detail: 90~degree bending and focussing of the beam.}
\label{f:trnsd3}
\end{figure}

After D2, the electrons travel in a horizontal beam pipe along the top 
of SK. 
Calibration holes reaching through the OD into the ID at regular intervals 
allow insertion of a vertical beam pipe of variable length. 
A 90~degree bending magnet, D3, bends the horizontal beam into this vertical 
beam pipe. 
Before and after D3, sets of quadrupole-magnets (Q-magnets) focus the beam 
onto the endcap of the vertical beam pipe in the tank (Fig.~\ref{f:trnsd3}). 

\begin{figure}
\begin{center}
\resizebox{4cm}{!}{\includegraphics{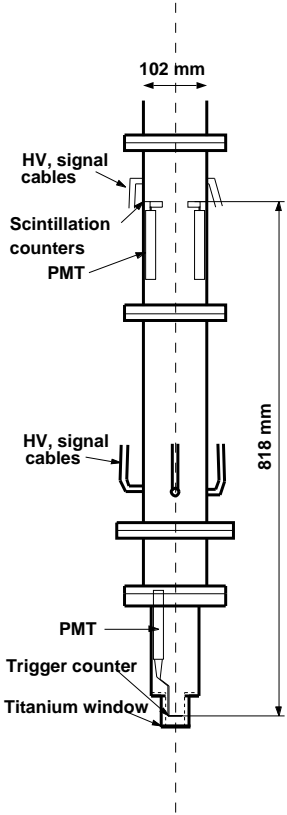}}
\end{center}
\caption{Beamline detail: Endcap}
\label{f:endcap}
\end{figure}

The endcap of the beam pipe is closed by a 100~$\mu$m thick titanium window 
of 3~cm diameter (Fig.~\ref{f:endcap}). 
In addition to separating the inside vacuum from outside water pressure of up 
to 4~atm, it must transmit electrons of energies down to 5~MeV with as little 
energy loss and multiple scattering as possible. 
The pointed shape that is approximated by cylinders of decreasing diameter 
towards the titanium exit window reduces shadowing by the beam pipe in the 
electrons' backward direction. 
A 1~mm thick, 24~mm diameter plastic scintillator (the ``trigger counter''),  
is mounted 17~mm above the titanium window and supplies a trigger signal. 
LINAC triggers are issued if a trigger counter hit coincides with a 
LINAC microwave pulse. 
Four scintillators of 1~cm thickness surround the beam 80~cm above the trigger 
counter. 
They are used when steering the beam onto the trigger counter. 

Steering magnets S1 and S2, installed after D1 and D2 respectively, are adjusted 
to keep the beam on the pipe axis. 
Since D1 and D2 take care of bending the beam out of and into the horizontal 
plane, the steering magnets need only be effective in the orthogonal direction. 
After D3, another set of magnets (XY-steering) installed right above the 
water level in the tank steers the beam in two orthogonal 
directions, independent of D3. 
These magnets are also used to scan the beam profile at the trigger counter. 
Measured beam profiles are of order 2~cm wide or less, well contained 
within the trigger counter. 

While fixed settings of D1 are used to control the beam momentum, all other 
magnet settings are tuned for minimal beam loss. 
Beam monitors before D2 and D3 can be inserted into the beam without breaking 
the vacuum in the beam pipe. 
Only the Q-magnet settings rely on a simulation of the beam optics. 
The validity of the Q-magnet settings is borne out by measurements with the 
scintillation counters in the endcap. 
The beam intensity is controlled by the electron gun. 
To ensure single electrons in detector, the beam intensity is reduced to 
$\sim$0.1 electrons per microwave pulse. 
Specifications of the magnets in the beam line are shown in Tab.~\ref{t:mag}.

\begin{table}
\begin{tabular}{|l|r|c|c|l|}  \hline
Name & Purpose & pole gap & Length & Magnetic field (for 16.3 MeV/c)\\ \hline

D1 & 15 degrees bending  & 2.5 cm & 7.5 cm & 1.3 KG \\
D2 & 15 degrees bending  & 2.5 cm & 7.5 cm & 1.3 KG \\
D3 & 90 degrees bending  & 2.5 cm & 39.3 cm & 2.0 KG \\
S1 & horizontal steering & 3.8 cm & 4.0 cm & 18G/A (typically 1.0 A) \\
S2 & horizontal steering & 3.8 cm & 4.0 cm & 18G/A (typically 1.0 A) \\
Q1 & doublet quadropole & 4 cm & 2$\times$10cm & 48G/cm/A (1.54A and 3.81A) \\
Q2 & doublet quadropole & 4 cm & 2$\times$10cm & 48G/cm/A (2.70A and 2.86A) \\
XY & steering in SK tank & 5.2 cm & 30 cm & 12.6G/A (typically 0.2 A) \\
\hline
\end{tabular}
\caption{Beamline magnet specifications.}
\label{t:mag}
\end{table}

A mu-metal shield inside the horizontal and vertical pipes protects the beam 
from external magnetic fields. 
Unfortunately this mu-metal shield is missing in the vicinity of elements 
like magnets, monitors and collimators in the beam pipe. 
Especially around the D3 magnet, there are many such elements, leaving the beam 
exposed. 
The net effect is a loss in beam intensity between the monitor 
M2 and the trigger counter. 
This loss gets worse for low beam momentum (up to 80\% for a 5~MeV beam), 
and has the potential of producing gamma-rays in the surrounding material. 
At the same time, the relative impact of gammas accompanying LINAC 
electrons in the tank would be largest at low momenta. 

The impact of such systematics can be evaluated from a comparison of different 
types of ``empty'' triggers. 
Since the timing of the momentum selected electrons is constrained within a 
few hundred nanoseconds relative to the 2~$\mu$s LINAC microwave pulse, 
a trigger can be set up independent of the trigger scintillator in the endcap. 
Ninety percent of the time this trigger will record ``empty'' events, where no 
electron is seen in the trigger counter. 
Taken in a separate run, this data set constitutes the so-called ``microwave 
triggers.'' 
Empty events for which detector readout is triggered by an external 100~Hz 
clock, independent of the LINAC, are taken in yet another run. 
Comparison of background rates in these two empty trigger runs, 
which are routinely taken after each LINAC run, 
no significant effect can be seen in the data. 
Differences between the respective background rates fluctuate around zero, and 
a conservative estimate for the systematic error of the absolute energy scale 
from beam correlated background is obtained 
by averaging the absolute values of these fluctuations, yielding 0.16\%. 

Positions in the detector that are accessible to the LINAC are restricted 
by the arrangement of calibration holes, which are situated with 2.12~m spacing 
(three PMT gridpoints along the top of the detector) along the radius that in SK 
coordinates is referred to as the x-axis. 
The vertical axis of the detector is the z-axis (see Fig.~\ref{f:gbl}). 
The origin is placed at the center of the ID. 
For structural reasons the calibration holes are offset from the radial 
x-axis by one PMT spacing (0.707~m). 
Only three of these holes were used in the current SK calibration. 
The z-coordinate of the endcap, where the electrons are 
delivered into the detector, is determined by the length of the vertical 
beam pipe. 
The necessity to reconfigure the beam pipe every time the LINAC position in 
the detector is changed automatically organizes LINAC data taking by 
position. 
For each position data are taken at seven different beam momenta between 
5.08 and 16.31 MeV/c. 
Currently all LINAC electrons exit the endcap in the -z direction. 

\section{Beam Energy Calibration}  \label{secBE}

The absolute energy of the beam is measured with a germanium detector.
After data are taken at a certain position, the D3 magnet 
is removed and the vertical beam pipe pulled out of the tank. 
To calibrate beam energy, the last section of the vertical beam 
pipe (containing the trigger counter) is connected directly 
to the horizontal beam pipe, so that it lies horizontally rather than 
hangs vertically as in the tank. 
The germanium detector is placed right after the titanium window, and 
D1 is set to the same value as for the measurement in SK.  
Ge calibration relates D1 magnet settings to beam energies. 

The germanium detector used in the calibration is a Seiko EG\&G Ortec 
GMX-35210-P, which has a germanium crystal of 57.5~mm diameter and 66.4~mm 
length.
The resolution of this germanium detector is 1.92~keV for the 1.33~MeV 
gamma-rays of $^{60}$Co. 
The germanium detector is connected to a Seiko EG\&G Multi-Channel-Analyzer 
7700. 

The germanium calibration system (crystal, electronics and 
multi-channel-analyzer (MCA)) is calibrated each time it is used. 
A variety of gamma-line sources, spanning an 
energy range from 0.662~MeV for a $^{137}$Cs source to 9.000~MeV from the 
Ni(n,$\gamma$)Ni reaction, establishes the relationship between energy and 
MCA channel. 
The linearity of this system calibration is found to be better than 1.5 keV 
over the 
whole energy range, corresponding to an error of less than 0.03\% in the 
momentum of a 5~MeV/c electron beam. 
The stability of the system is also excellent: without recalibration, the 
systematic error would still be less than 0.1\%. 

\begin{figure}
\begin{center}
\resizebox{8cm}{!}{\includegraphics{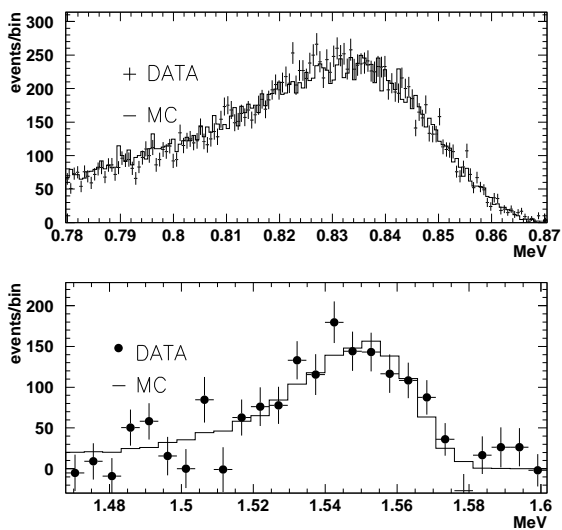}}
\end{center}
\caption{Data and MC for $^{207}$Bi internal conversion electrons. 
Both data and MC include 
energy loss in a 500~$\mu$m Be window and a 41~$\mu$m passive layer in 
the Ge crystal. The top figure is for 975.7~keV electrons, 
and the bottom for 1682.2~keV.}
\label{f:aircore}
\end{figure}

However, Ge detectors respond slightly differently to electrons and 
gamma-rays.
Incident electrons lose energy in the crystal's beryllium entrance window 
and in a passive layer at the crystal surface before they reach the active 
volume, 
while gamma-ray conversions mostly take place inside the active volume. 
To measure this initial energy loss, the Ge detector was taken to an 
air-core beta spectrometer at the Tanashi-branch of KEK. 
At the spectrometer, internal conversion electrons of 975.7 and 1682.2~keV 
from a $^{207}$Bi source were injected into the crystal. 
The average energy loss measured for these lines was 143.5 and 132.3~keV, 
respectively. 
Fig.~\ref{f:aircore} shows the background subtracted spectrometer data, which 
are matched by Monte Carlo (MC) simulation under the assumption of a 41~$\mu$m 
passive layer 
behind the 500~$\mu$m beryllium entrance window to the crystal. 
A scan across the entrance window did not reveal any inhomogeneities. 

\begin{figure}
\begin{center}
\resizebox{14cm}{!}{\includegraphics{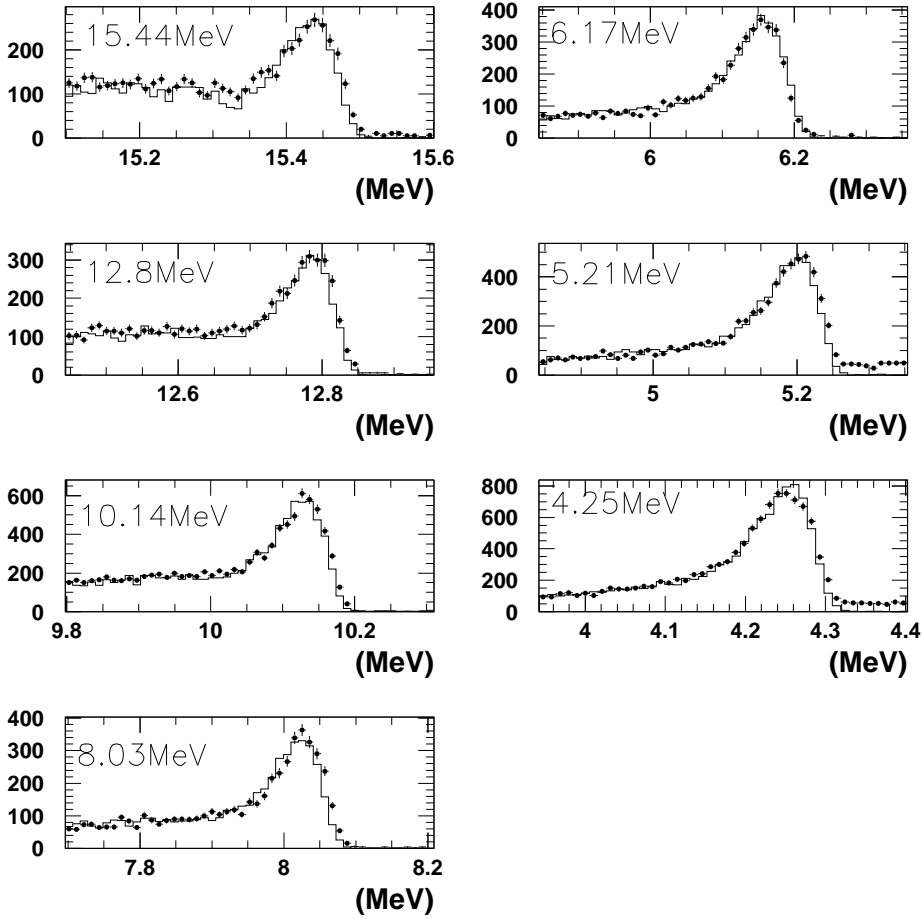}}
\end{center}
\caption{Ge calibration data (points) and MC simulation (histogram) for 
LINAC beams of various energies (see Tab.~\protect{\ref{t:ge}}). 
One bin corresponds to 10~keV.}
\label{f:Gedat}
\end{figure}

A MC simulation is used to evaluate the impact of energy loss and multiple 
scattering in the trigger counter, the titanium window, the beryllium window 
and the passive layer of the Ge crystal. 
The vacuum in the beam pipe is better than 10$^{-4}$~Torr, so energy loss in 
the rest gas can be neglected. 
In Fig.~\ref{f:Gedat}, spectra recorded in the MCA are displayed for various 
beam momenta, as selected by the setting of the D1 magnet. 
Tails towards lower energies are due to electrons that are not contained in 
the Ge crystal. 
The simulation reproduces the width of the spectra to better than 10~keV.
The results of the beam momentum calibration are summarized in Tab.~\ref{t:ge}. 
Five Ge calibrations were done during the time the LINAC data for SK 
calibration were collected. 
Like the SK calibration data, Ge data were taken at three different x-positions. 
The reproducibility of the relationship between D1 magnet setting and beam 
energy is better than 20~keV. 

\begin{table}
\begin{center}
\begin{tabular}{|c|c|c|c|}  \hline
D1 current  & beam momentum & Ge energy & in-tank energy \\
(A)         & (MeV/c)       & (MeV)     & (MeV) \\ \hline
1.8         & 5.08          & 4.25      & 4.89 \\
2.15        & 6.03          & 5.21      & 5.84 \\
2.5         & 7.00          & 6.17      & 6.79 \\
3.2         & 8.86          & 8.03      & 8.67 \\
4.0         & 10.99         & 10.14     & 10.78 \\
5.0         & 13.65         & 12.80     & 13.44 \\
6.0         & 16.31         & 15.44     & 16.09 \\ \hline 
\end{tabular}
\end{center}
\caption{D1 setting and associated beam momentum. 
The third column gives the energy 
measured in the Ge calibration system. The last column lists the total energy 
of the electrons after leaving the beampipe. }
\label{t:ge}
\end{table}

\section{Super-Kamiokande Calibration}  \label{secDA}

After installation of the LINAC in 1996, testing and commissioning of the 
various components proceeded through the summer of 1997. 
Systematic LINAC data taking in SK started in September 1997. 
High quality data sets were obtained for eight different positions in the ID 
(Tab.~\ref{t:dpos} and Fig.~\ref{f:gbl}). 
With a maximal repetition rate of 66~Hz and only 10\% of pulses 
actually delivering an electron, it takes about two hours to collect the 
necessary statistics for one energy setting. 

\begin{table}
\begin{center}
\begin{tabular}{|c|c|c|c|}  \hline
Position & X (m) & Y (m) & Z (m) \\ \hline
A & -3.88 & -0.71 & 12.28 \\
B & -3.88 & -0.71 &  0.27 \\ 
C & -8.13 & -0.71 &  12.28 \\
D & -8.13 & -0.71 & 0.27 \\
E & -12.37 & -0.71 & 12.28 \\
F & -12.37 & -0.71 & 0.27 \\ 
G & -3.88 & -0.71 & -11.73 \\
H & -12.37 & -0.71 & -11.73 \\ \hline
\end{tabular}
\end{center}
\caption{List of positions where LINAC data were taken}
\label{t:dpos}
\end{table}

\begin{figure}
\resizebox{12cm}{!}{\includegraphics{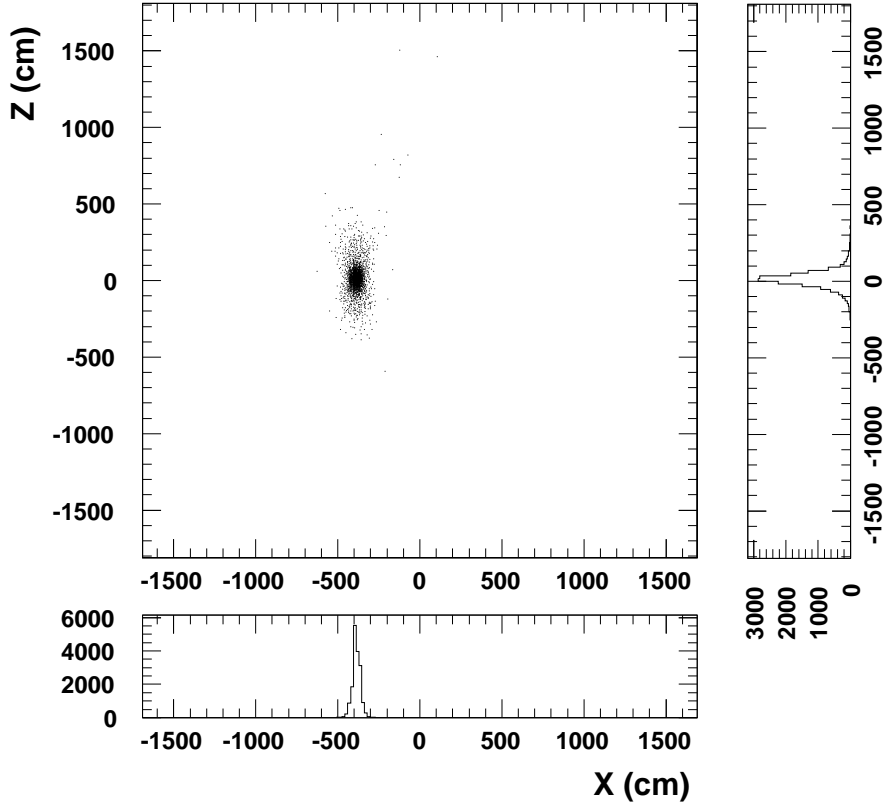}}
\caption{Vertex position distribution of LINAC data taken at 
(x,z)=(-4m,0m), beam momentum 16.31 MeV/c. Projections of the scatter plot 
are shown to the right and underneath. Scatter plot limits correspond to 
the limits of the ID.}
\label{f:tvtx}
\end{figure}

LINAC data are reconstructed using the standard solar neutrino 
analysis chain (see Ref.\cite{ref:sksol}). 
A resulting typical two-dimensional projection of the vertex distribution 
onto the x,z-plane (SK coordinates) for LINAC trigger events is shown in 
Fig.~\ref{f:tvtx}. 
Also shown are the corresponding projections onto the x and z axes. 

\begin{figure}
\begin{center}
\resizebox{10cm}{!}{\includegraphics{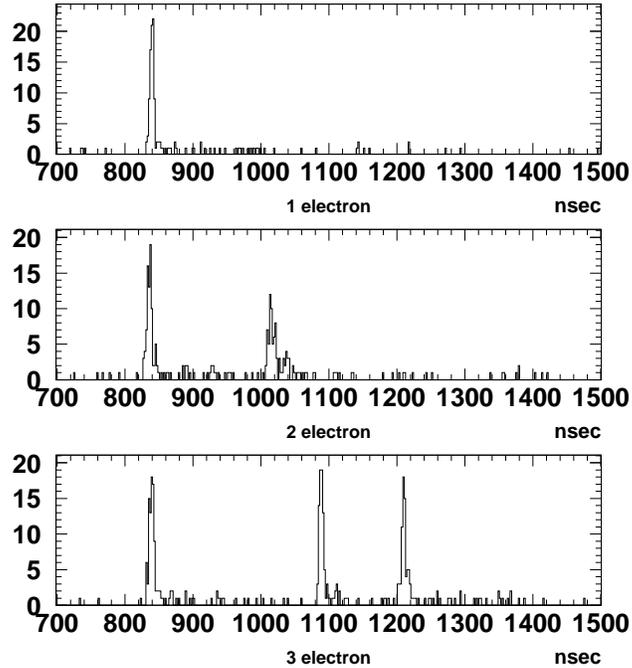}}
\end{center}
\caption{Examples for TOF subtracted timing information for LINAC events 
containing one, two and three electrons.}
\label{f:npeak}
\end{figure}

To select the calibration data from this data set, one additional cut is 
applied to these LINAC trigger data to reject events with multiple electrons. 
If the timing information for the event is corrected for the time of flight 
(TOF) from the end of the beampipe, electrons that left the beampipe at times 
different by only a few tens of nanoseconds can be clearly separated. 
Examples for such TOF subtracted timing distributions are shown in 
Fig.~\ref{f:npeak}. 
Events are rejected if multiple peaks of more than 30\% of the expected signal 
are found more than 30~ns apart in a single event. 
About 5\% of the LINAC trigger events are rejected by this cut. 

The resulting set of LINAC calibration data for the various positions and 
energies is used in two ways. 
First, it is used to tune parameters in the detector simulation; second, it 
is used to evaluate systematic errors from the remaining discrepancies. 
While the LINAC provides samples of data for limited sets of 
positions and energies and (until now) only one fixed direction, a single 
MC description covers the whole range of solar neutrino events.  
LINAC calibration data and MC simulation output will hereafter be 
referred to as LINAC and MC. 
MC simulation is based on GEANT version 3.21\cite{ref:geant}. 
Calibration equipment is included in this simulation. 
All MC shown in the figures uses the current best set of tuned parameters. 
New insights in details of the detector response may change this MC 
description and the derived parameters.   

\begin{figure}
\begin{center}
\resizebox{16cm}{!}{\includegraphics{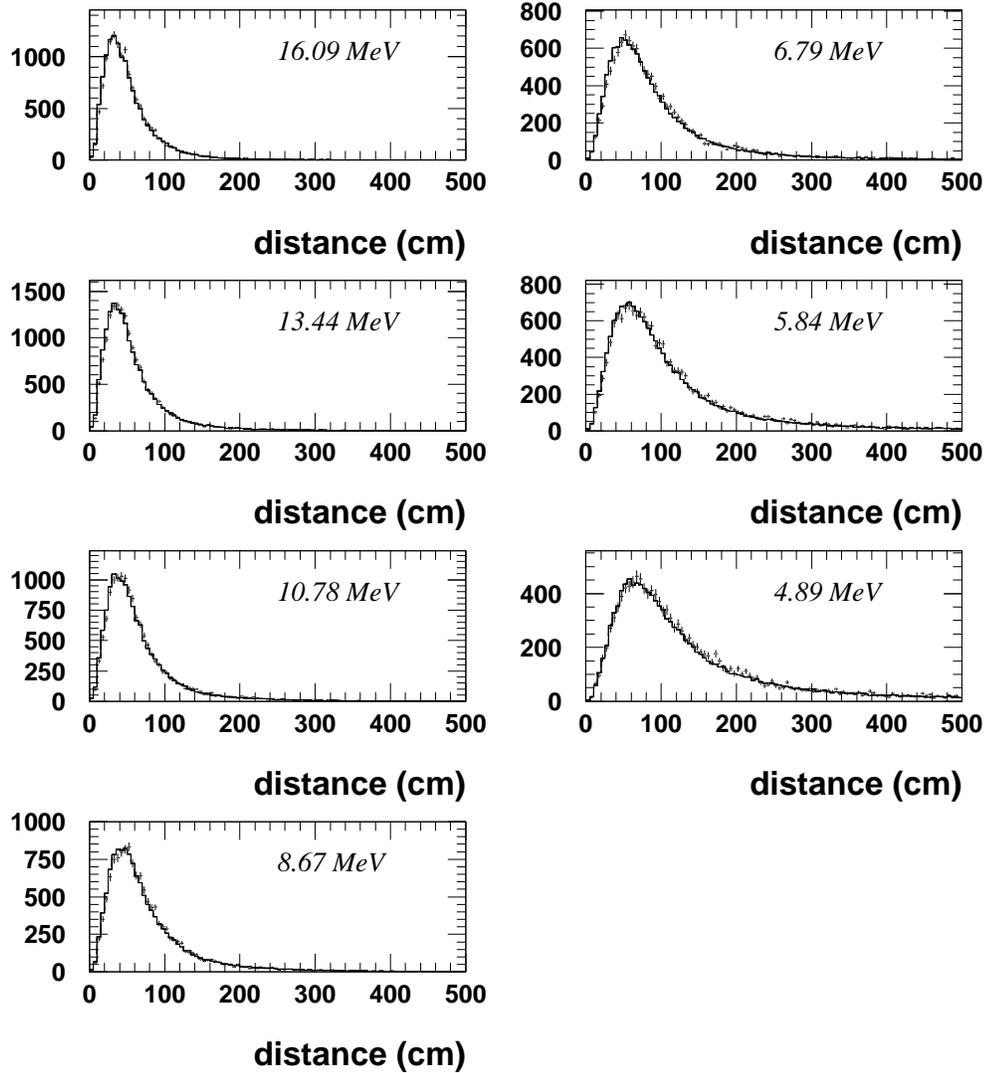}}
\end{center}
\caption{Vertex distributions for (x,z)=(-12m,+12m). 
Histogram is MC, points LINAC.}
\label{f:reVTX}
\end{figure}

\begin{figure}
\begin{center}
\resizebox{65mm}{!}{\includegraphics{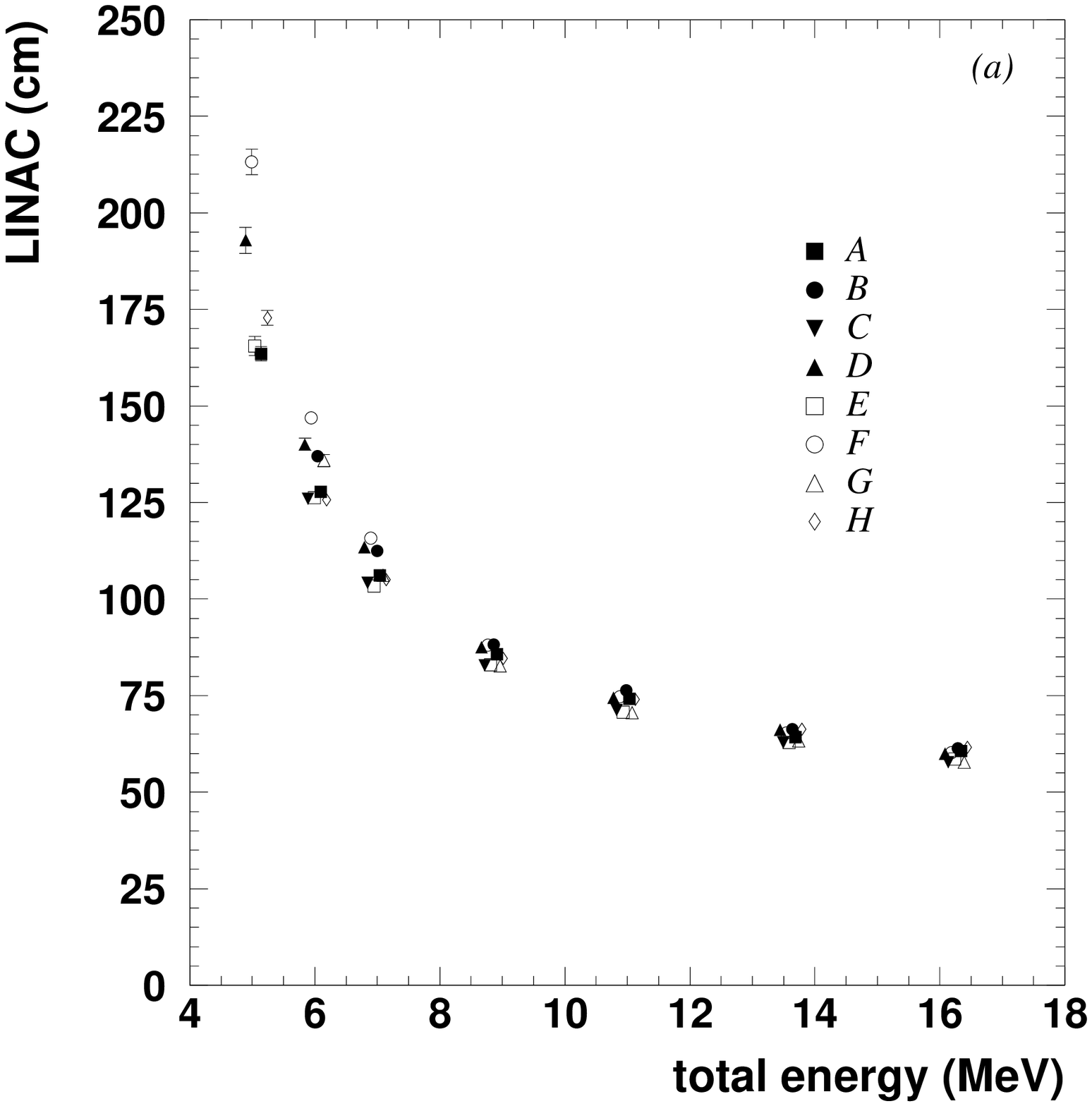}}
\resizebox{65mm}{!}{\includegraphics{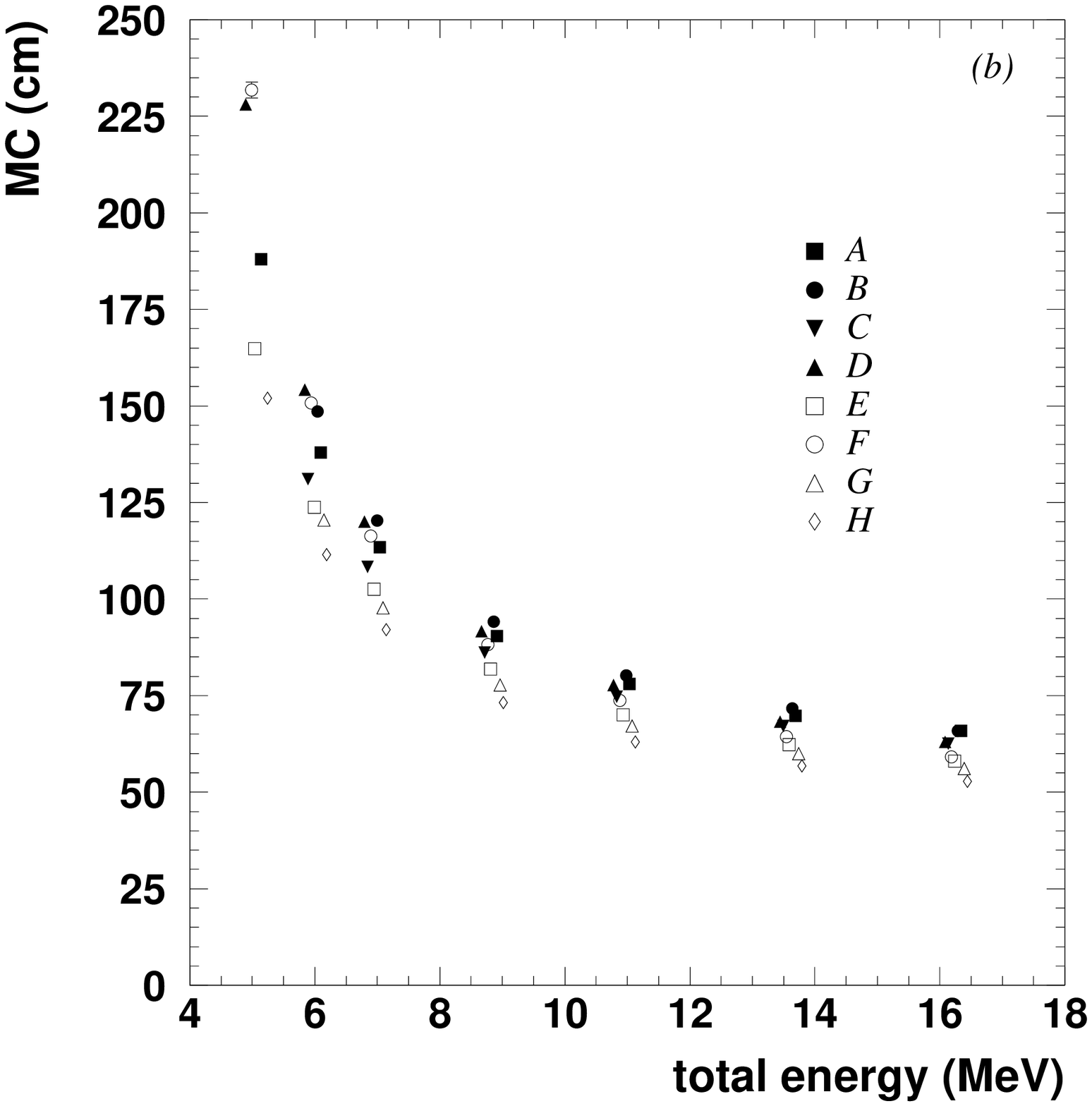}}
\resizebox{65mm}{!}{\includegraphics{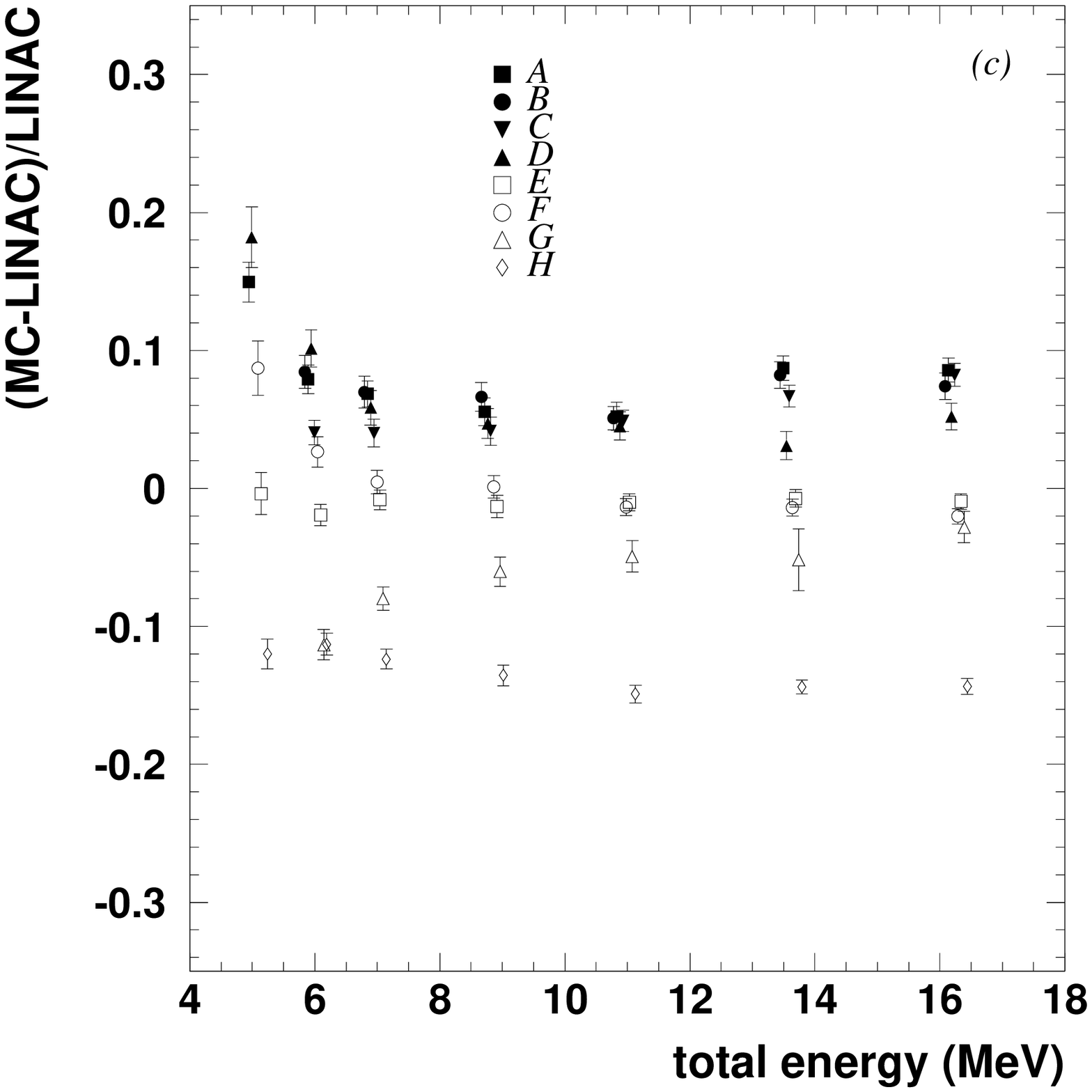}}
\resizebox{65mm}{!}{\includegraphics{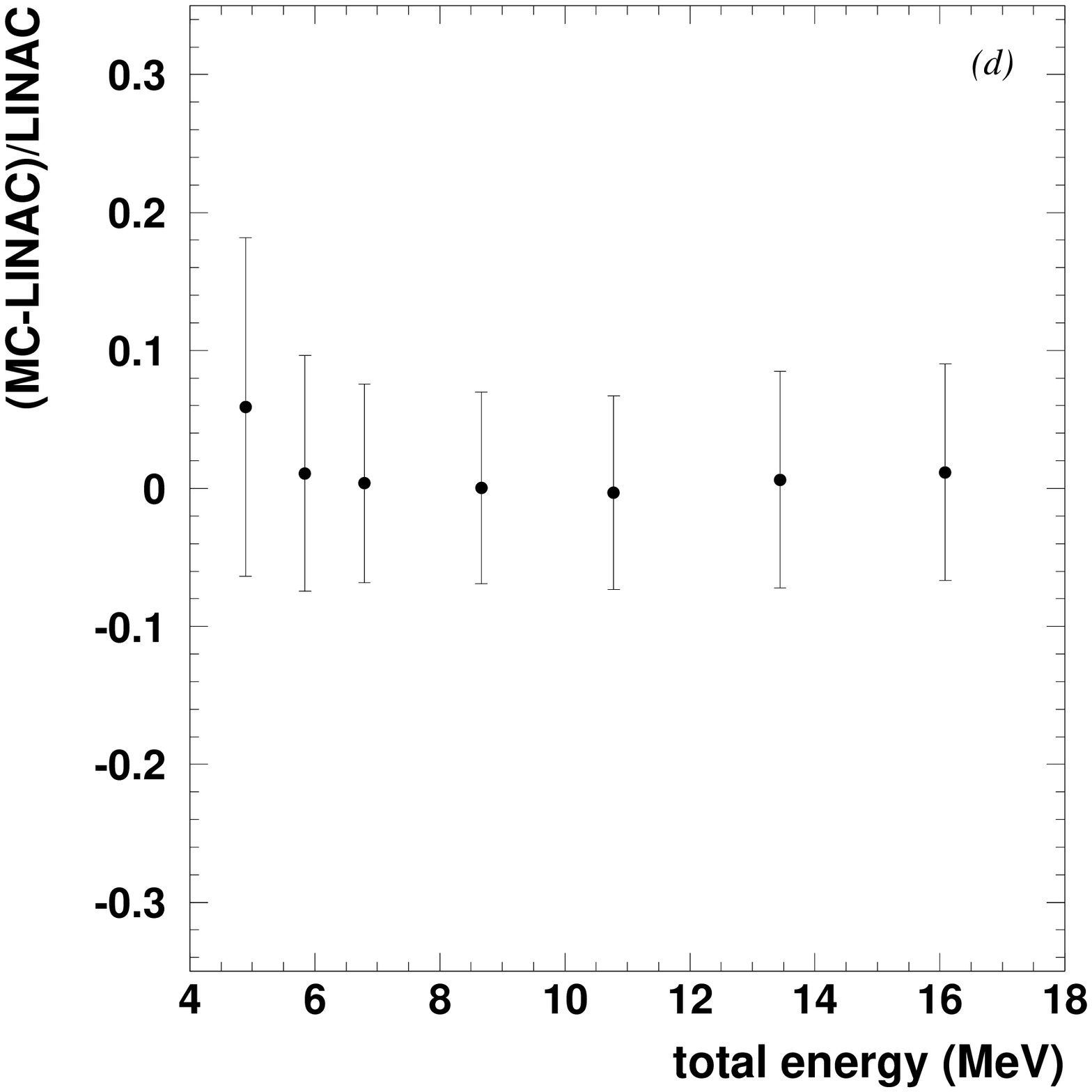}}
\end{center}
\caption{Vertex position resolutions of (a) LINAC and (b) MC. 
A--H are defined in Tab.~\protect{\ref{t:dpos}}.
(c) shows relative differences for all positions over momentum, (d) averages 
over position. Errors in (a)-(c) are statistical, while in (d) it is the RMS 
of the spread over positions in (c).}
\label{f:reVRES}
\end{figure}

The PMT timing resolution for single photo-electrons in the MC is tuned to 
reproduce the vertex resolution throughout the detector. 
The value of 2.4~ns derived here is in agreement with independent 
measurements on the SK PMTs reported in a previous NIM paper\cite{ref:tubes}. 
An example of MC and LINAC vertex distributions in terms of distance from the 
beampipe end position to the reconstructed vertex is shown in 
Fig.~\ref{f:reVTX}. 
Fig.~\ref{f:reVRES} shows the vertex resolution for all positions as a 
function of energy and the relative differences between MC and LINAC. 
Vertex resolution is defined as the radius of the sphere around the beampipe 
end position that contains 68\% of the reconstructed vertices. 
Errors in the position averaged diagram reflect the spread (RMS)
of values deduced for the individual LINAC positions.   

\begin{figure}
\begin{center}
\resizebox{16cm}{!}{\includegraphics{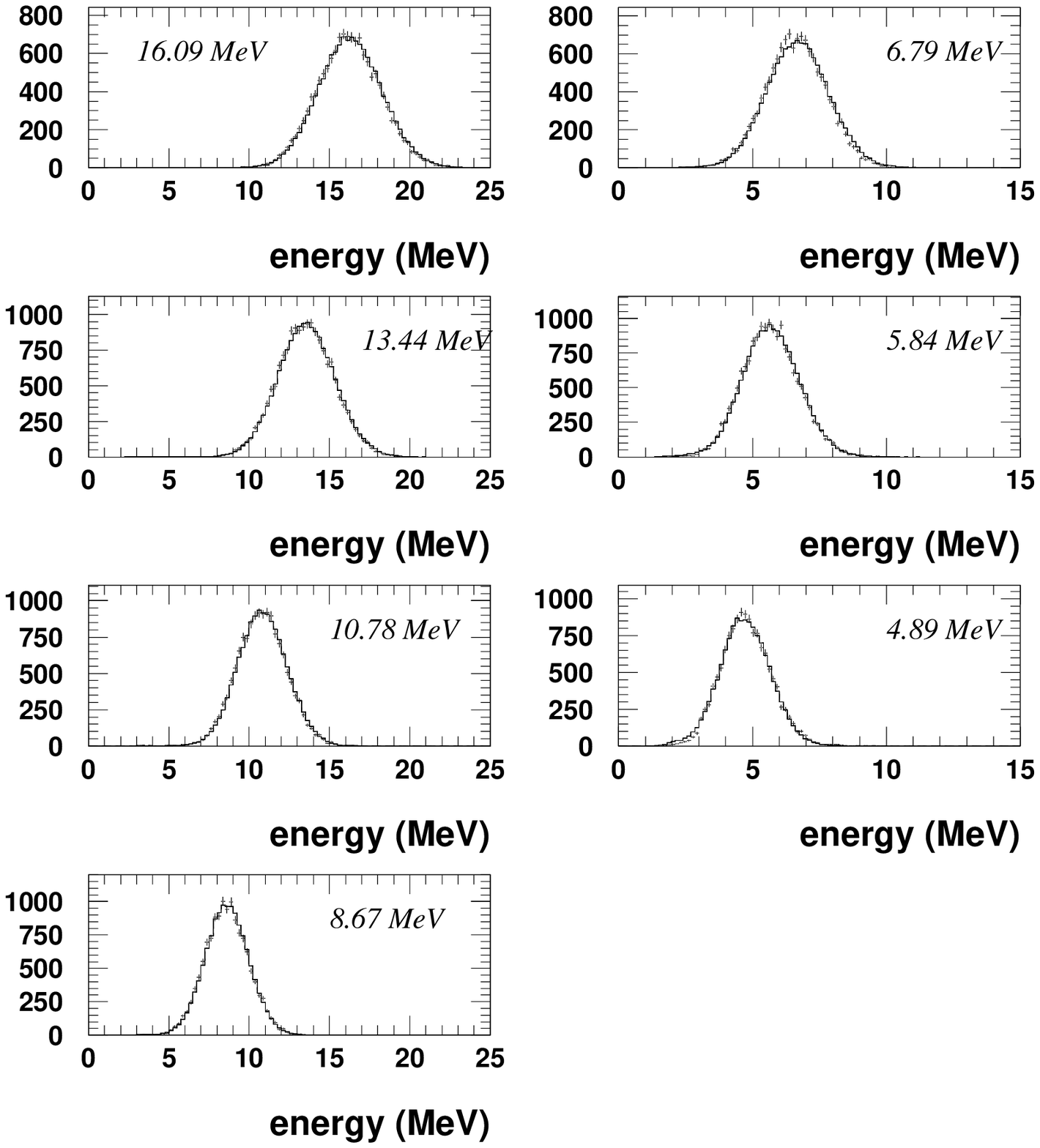}}
\end{center}
\caption{LINAC (crosses) and MC (histogram) energy distributions for 
\mbox{(x,z)=(-12m,+12m)}.}
\label{f:reET}
\end{figure}

Many MC parameters have influence on the energy scale. 
Its position dependence is mostly affected by Cherenkov light attenuation, 
while the overall scale is adjusted by the PMT collection efficiency. 
At short wavelengths, the attenuation length is limited by scattering. 
MC tuning fixes it at 59.4~m for 380~nm. 
The same result is obtained from a direct measurement in the SK tank, 
yielding 59.4$\pm$1.6~m at this wavelength\footnote{For wavelengths longer 
than 400~nm, the attenuation length becomes longer than 75~m, until absorption 
starts beyond $\sim$430~nm. Most of the Cherenkov light is 
detected near 380~nm.}. 
PMT quantum efficiency and the reflectivities of various detector materials 
are put into the simulation from direct measurements. 
MC and LINAC reconstructed energy distributions overlaid for one position in 
the tank are shown in Fig.~\ref{f:reET}. 

\begin{figure}
\begin{center}
\resizebox{65mm}{!}{\includegraphics{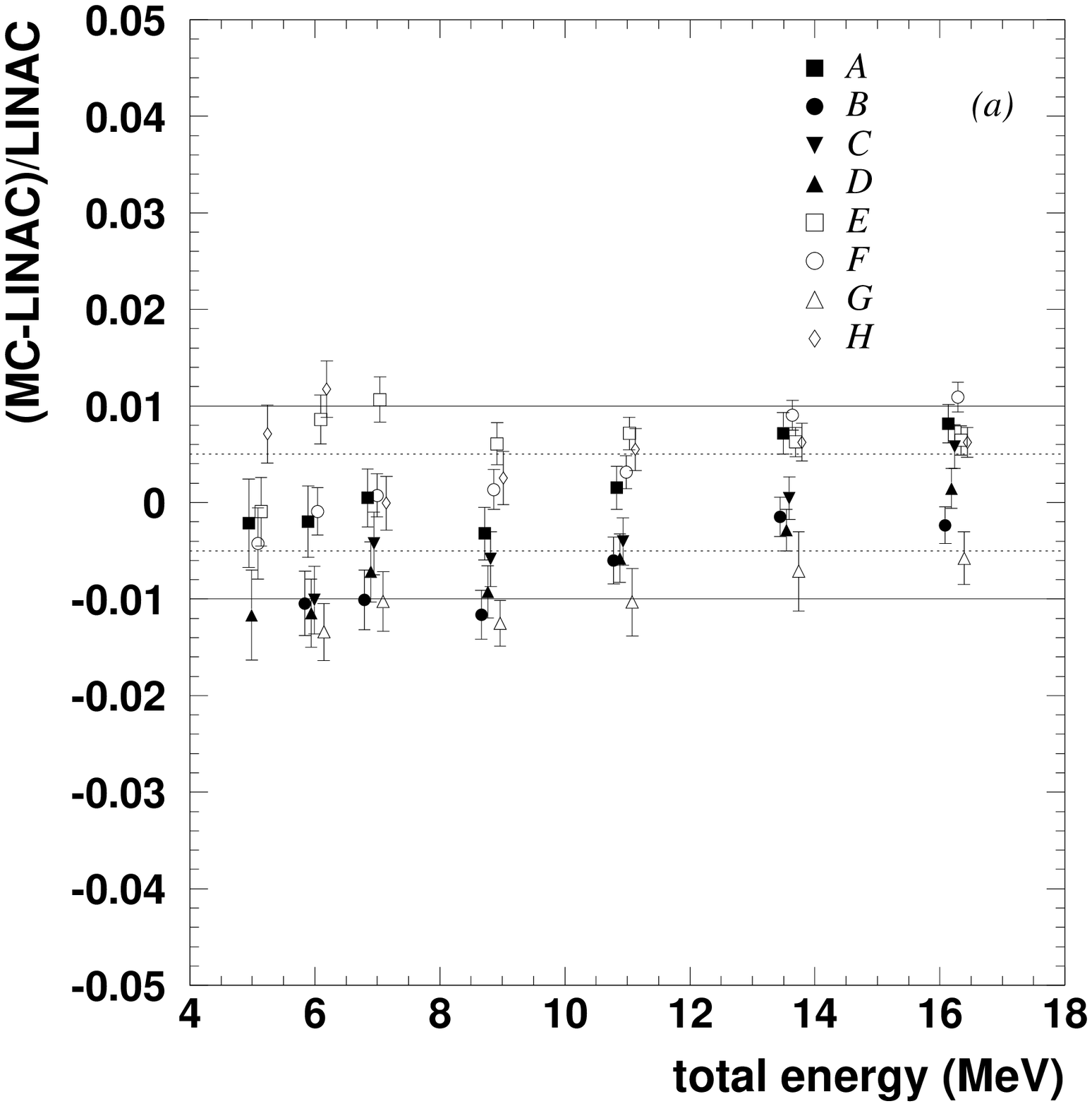}}
\resizebox{65mm}{!}{\includegraphics{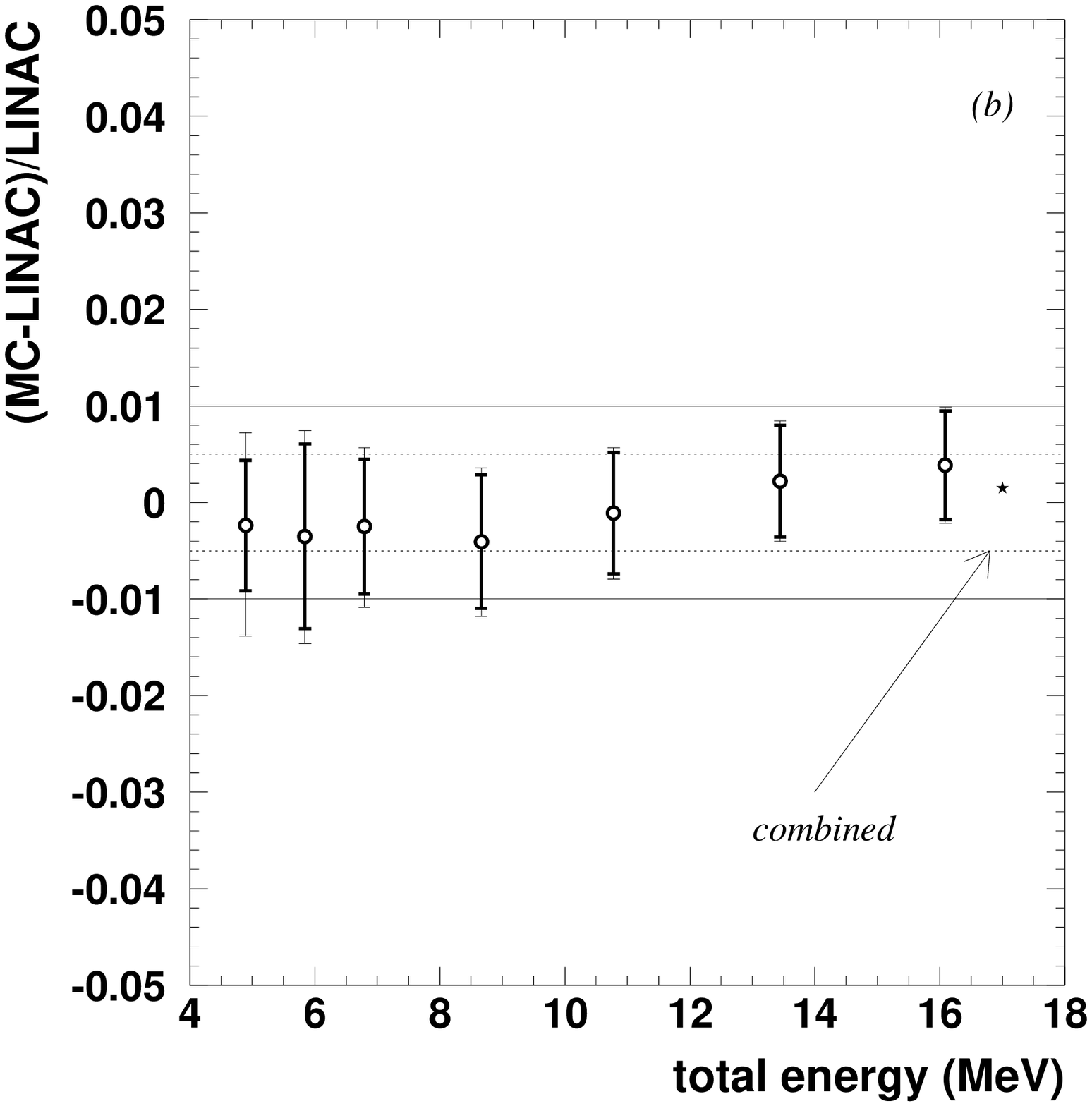}}
\end{center}
\caption{Comparison of absolute energy scales for LINAC and MC. 
A--H are defined in Tab.~\protect{\ref{t:dpos}}. 
(a) has only statistical errors. For the position averages shown in (b), 
the inner error is the RMS of the spread over position while the outer one is 
the systematic error. 
Dotted and solid lines show $\pm$0.005 and $\pm$0.01 in (MC-LINAC)/LINAC.
The last point in (b) represents the total average over all 
positions and beam energies.}
\label{f:reESCL}
\end{figure}

Fig.~\ref{f:reESCL} compares absolute energy scales as a function of 
both energy and position. 
For this comparison, MC and LINAC distributions, like the ones shown in 
Fig.~\ref{f:reET}, are reduced to peak value and width by a Gaussian fit 
around the center. 
In Fig.~\ref{f:reESCL}b the systematic error of the position averaged scale 
shift increases toward lower energies, but the central value always stays 
within $\pm$0.5\%.  
This increase in error is due to uncertainties in the reflectivity of the 
beampipe endcap's materials (see below). 
The statistical errors in this figure represent the spread (RMS) of the 
central values for the different positions. 

\begin{figure}
\begin{center}
\resizebox{65mm}{!}{\includegraphics{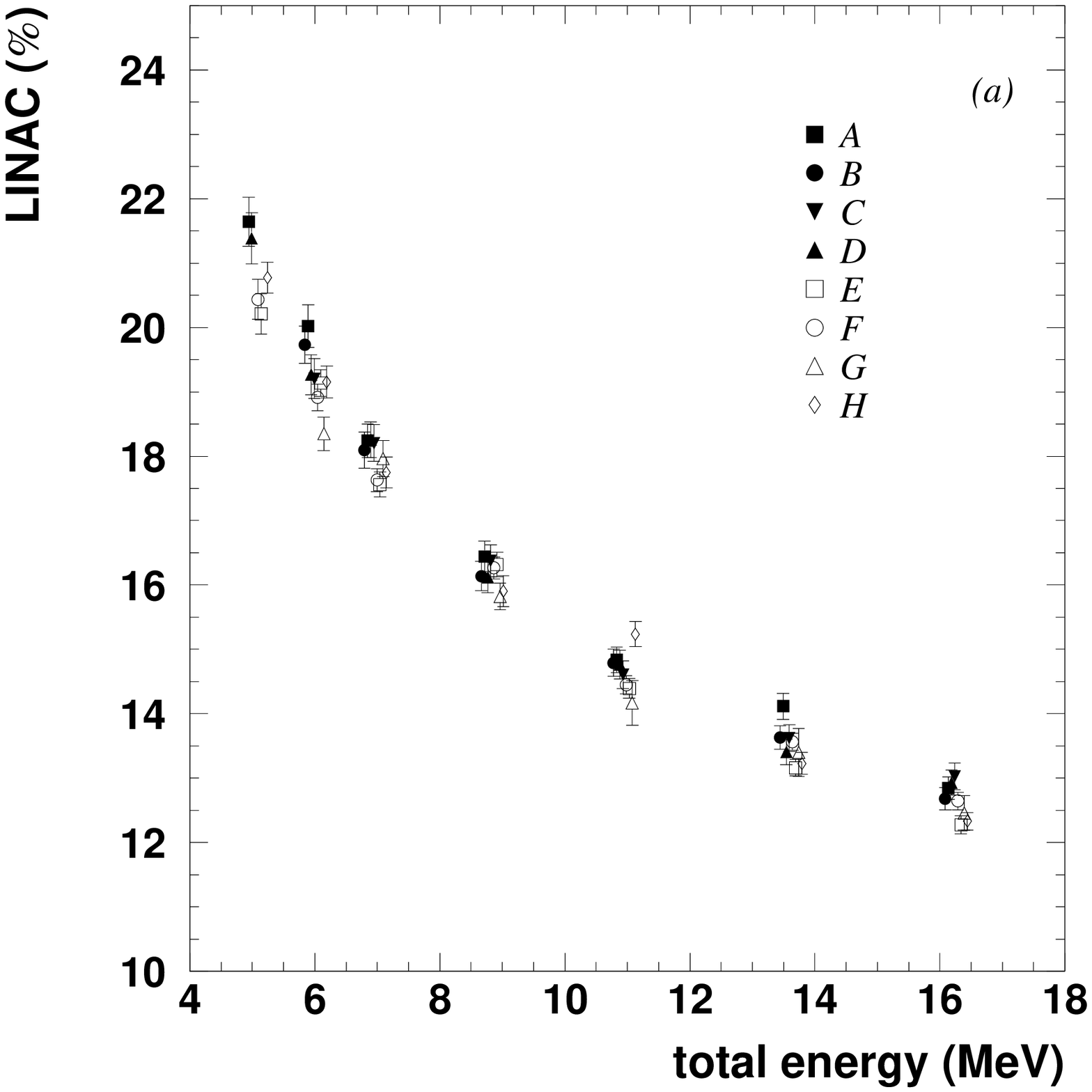}}
\resizebox{65mm}{!}{\includegraphics{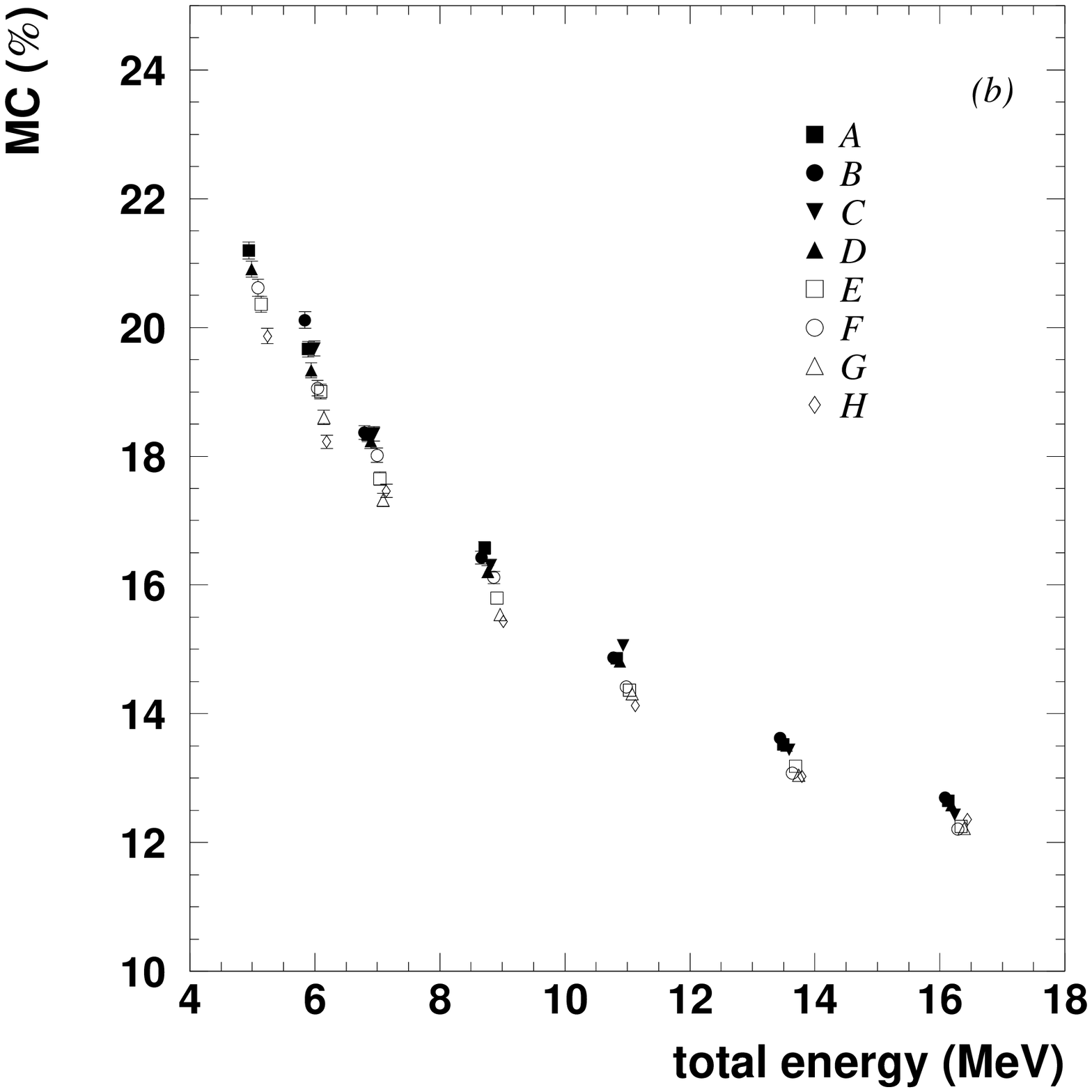}}
\resizebox{65mm}{!}{\includegraphics{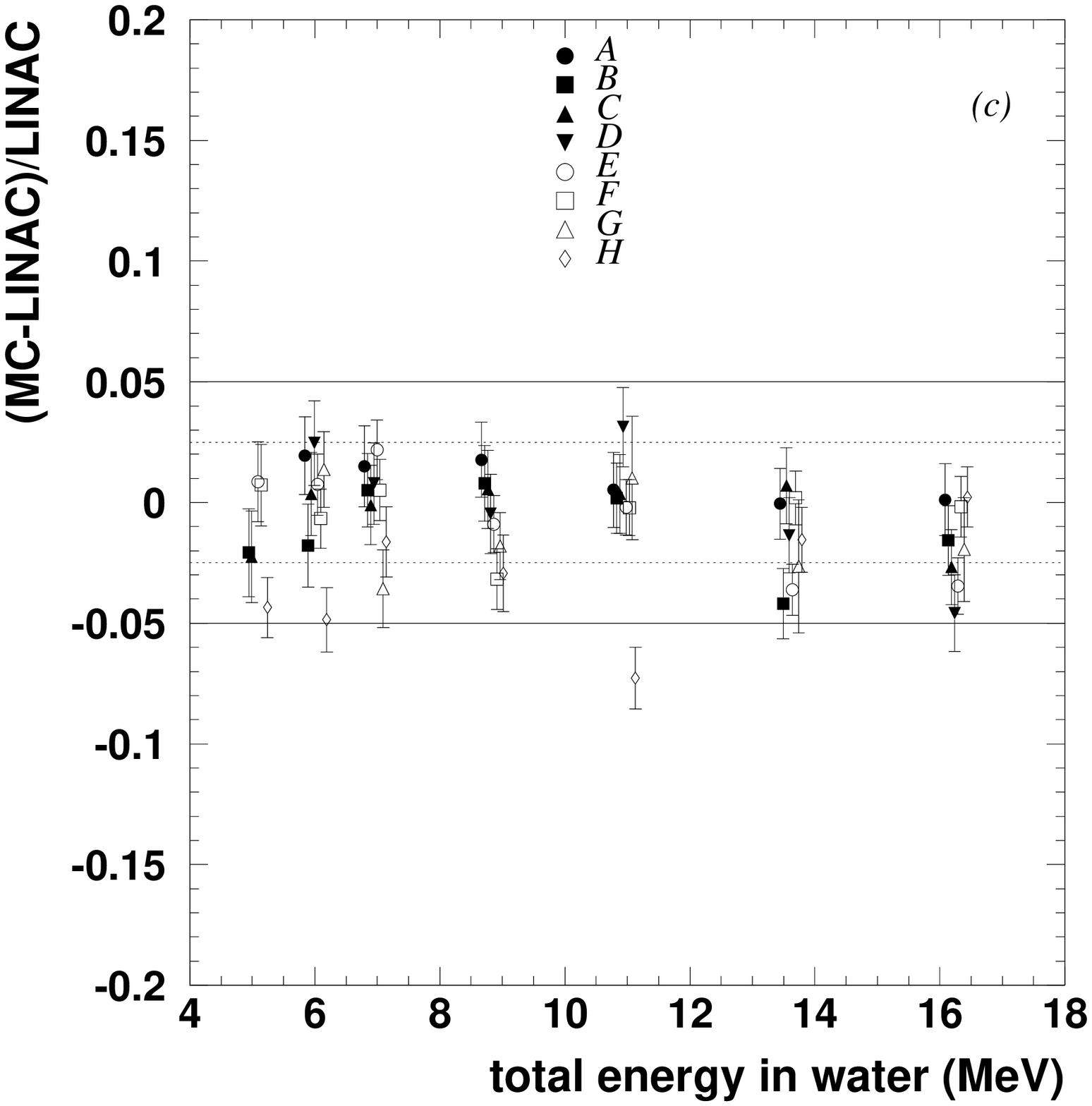}}
\resizebox{65mm}{!}{\includegraphics{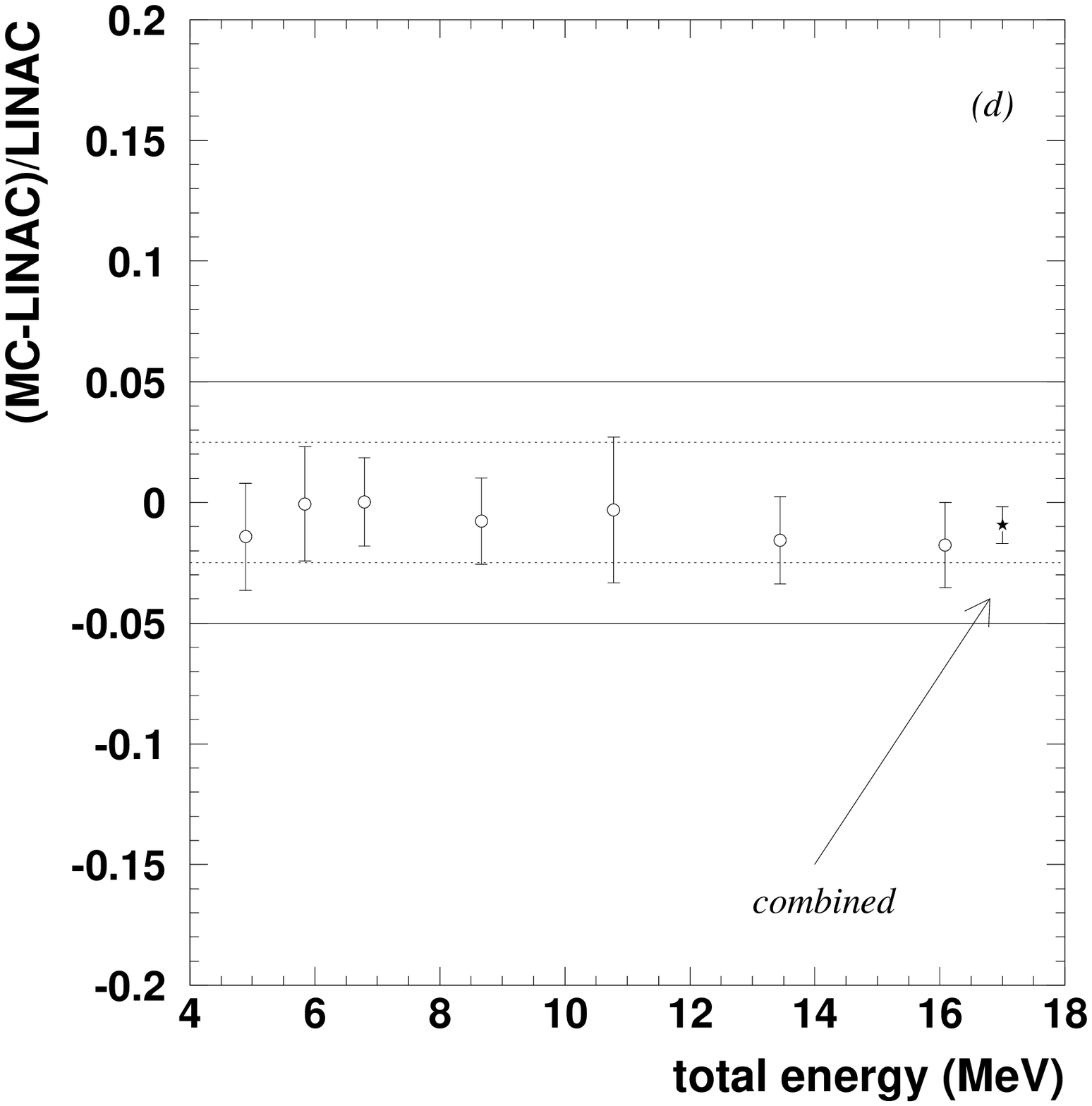}}
\end{center}
\caption{Energy dependence of energy resolution of (a) LINAC and (b) MC,
(c) relative difference between (a) and (b). (d) shows the position averages, 
and its last point the total average. Errors as in 
Fig.~\protect{\ref{f:reVRES}}. A--H are defined in Tab.~\protect{\ref{t:dpos}}. 
Solid and dotted lines are $\pm$0.05 and $\pm$0.025 in
(MC-LINAC)/LINAC.} 
\label{f:reERES}
\end{figure}

Energy dependence of the energy resolution for MC and LINAC is compared in 
Fig.~\ref{f:reERES}. 
The energy resolution is not tuned directly. 
It is defined as the width of the Gaussian fit divided by its peak value. 
The current MC reproduces the experimental distributions to within 2\%. 

Estimates for the systematic error of the absolute energy scale 
are given in Tab.~\ref{t:syserr}.
With uncertainties of the energy loss in the beam pipe endcap estimated to be 
2~keV, and 1.5~keV from the calibration of the germanium detector, the 
reproducibility of the momentum selection in D1 with 20~keV dominates the 
uncertainty in the LINAC beam momentum. 
Beam correlated background may contribute up to 0.16\% 
(see section \ref{secEL}). 

The most serious difficulty is the reflectivity of the endcap. 
At 5~MeV, about 5\% of the Cherenkov photons emitted by an electron leaving 
the beampipe will hit the endcap (see column 2 of Tab.~\ref{t:syserr}). 
Recent tests revealed that there is some danger for a bubble of air to be 
trapped within the rim closing the seal of the titanium window. 
Thus, although the reflectivity of the steel beampipe and the titanium window 
were measured, the presence of a bubble of air of unknown size in front of it 
would change the situation significantly. 
MC estimates for the extreme cases of no air and maximum bubble size are 
currently used to obtain a conservative estimate of this uncertainty. 
The numbers are given in column 3 of Tab.~\ref{t:syserr}. 

A cross-check validating the absolute energy scale established in the LINAC 
calibration comes from $^{16}$N decays. 
$^{16}$N is produced throughout the detector by stopping cosmic ray muons, 
providing a data sample free of the mechanical constraints of the LINAC. 
This data sample also is well reproduced by the MC simulation. 

\begin{table}
  \begin{tabular}{|c|c|c|c|}  \hline
    beam momentum (MeV/c) & fraction hitting & error due to  & total systematic error \\ 
     & beam pipe & reflectivity &  \\  \hline
    5.08  & 4.7\%  & $\pm$0.68\% & $\pm$0.81\% \\
    6.03  & 3.3\%  & $\pm$0.40\% & $\pm$0.55\% \\
    7.00  & 2.2\%  & $\pm$0.28\% & $\pm$0.44\% \\
    8.86  & 1.3\%  & $\pm$0.18\% & $\pm$0.33\%\\
    10.99 & 0.88\%  & $\pm$0.11\% & $\pm$0.27\% \\
    13.65 & 0.67\% & $\pm$0.08\% & $\pm$0.24\% \\
    16.31 & 0.51\%  & $\pm$0.06\% & $\pm$0.21\% \\ \hline 
  \end{tabular}
  \caption{Systematic errors at the various LINAC momenta. 
The first column shows beam momentum, the second the fraction of
    Cherenkov photons hitting the beam pipe, the third the systematic error due
    to uncertainty of the reflectivity of the endcap, and the last 
    the resulting total systematic error in the derived absolute energy scale.}
  \label{t:syserr}
\end{table}

\begin{figure}
\begin{center}
\resizebox{16cm}{!}{\includegraphics{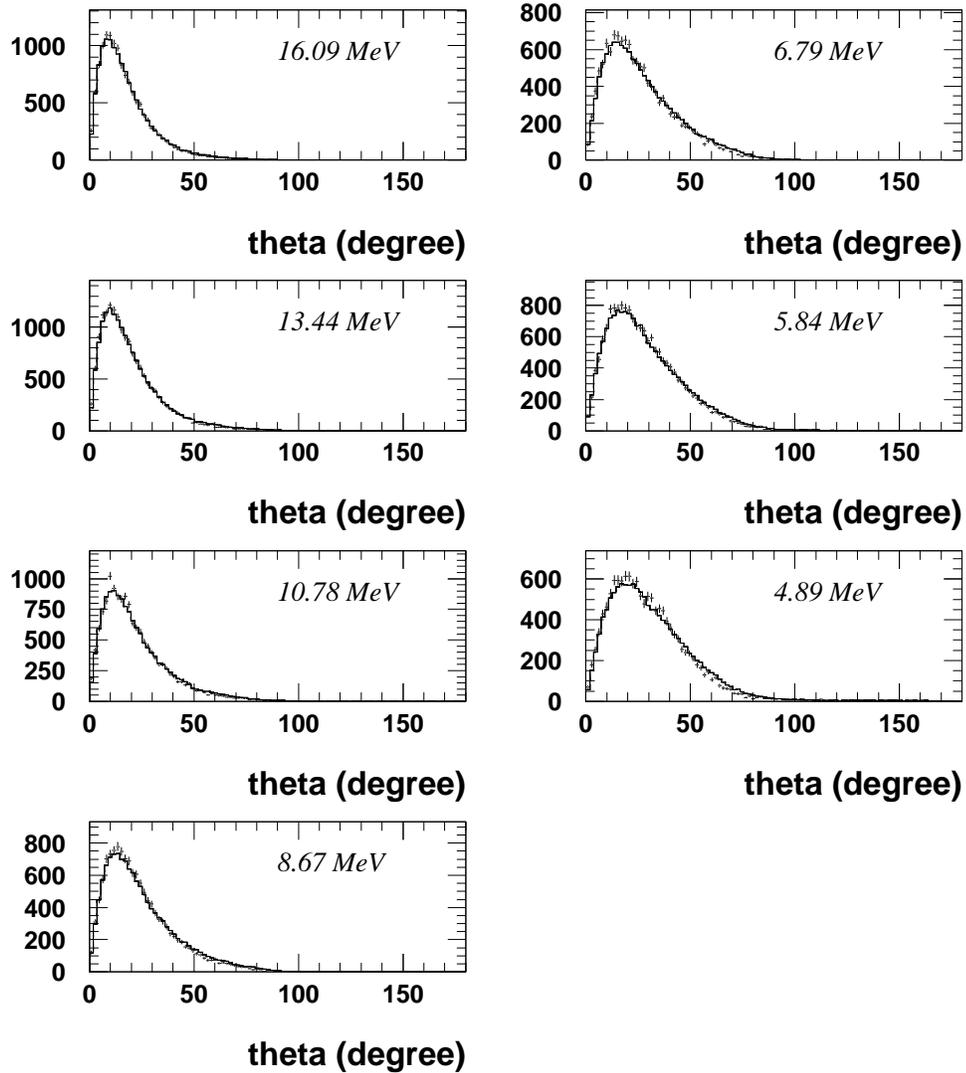}}
\end{center}
\caption{Angular distributions for (x,z)=(-12m,+12m). 
Histogram is MC, points LINAC.}
\label{f:reANG}
\end{figure}

\begin{figure}
\begin{center}
\resizebox{65mm}{!}{\includegraphics{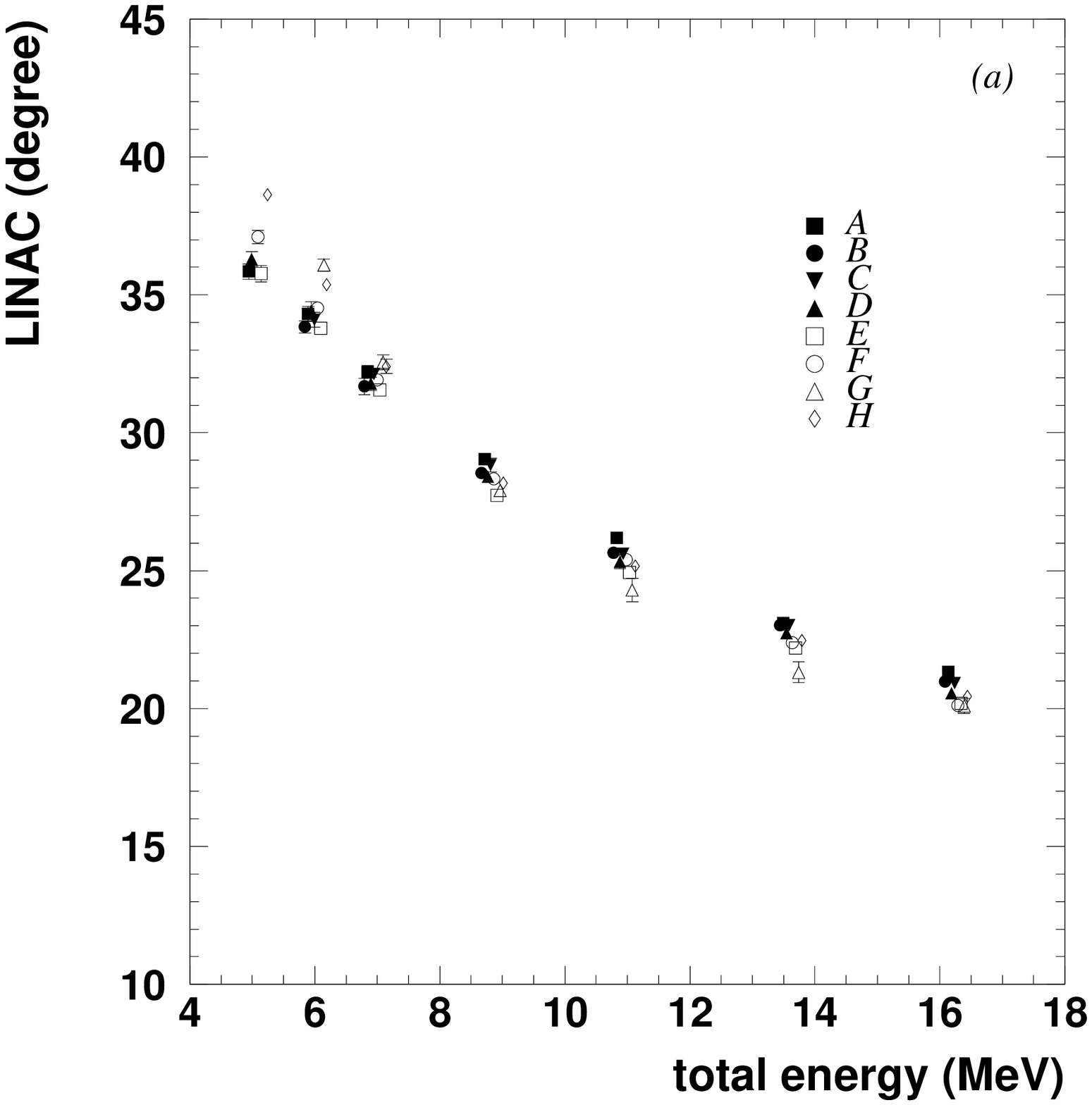}}
\resizebox{65mm}{!}{\includegraphics{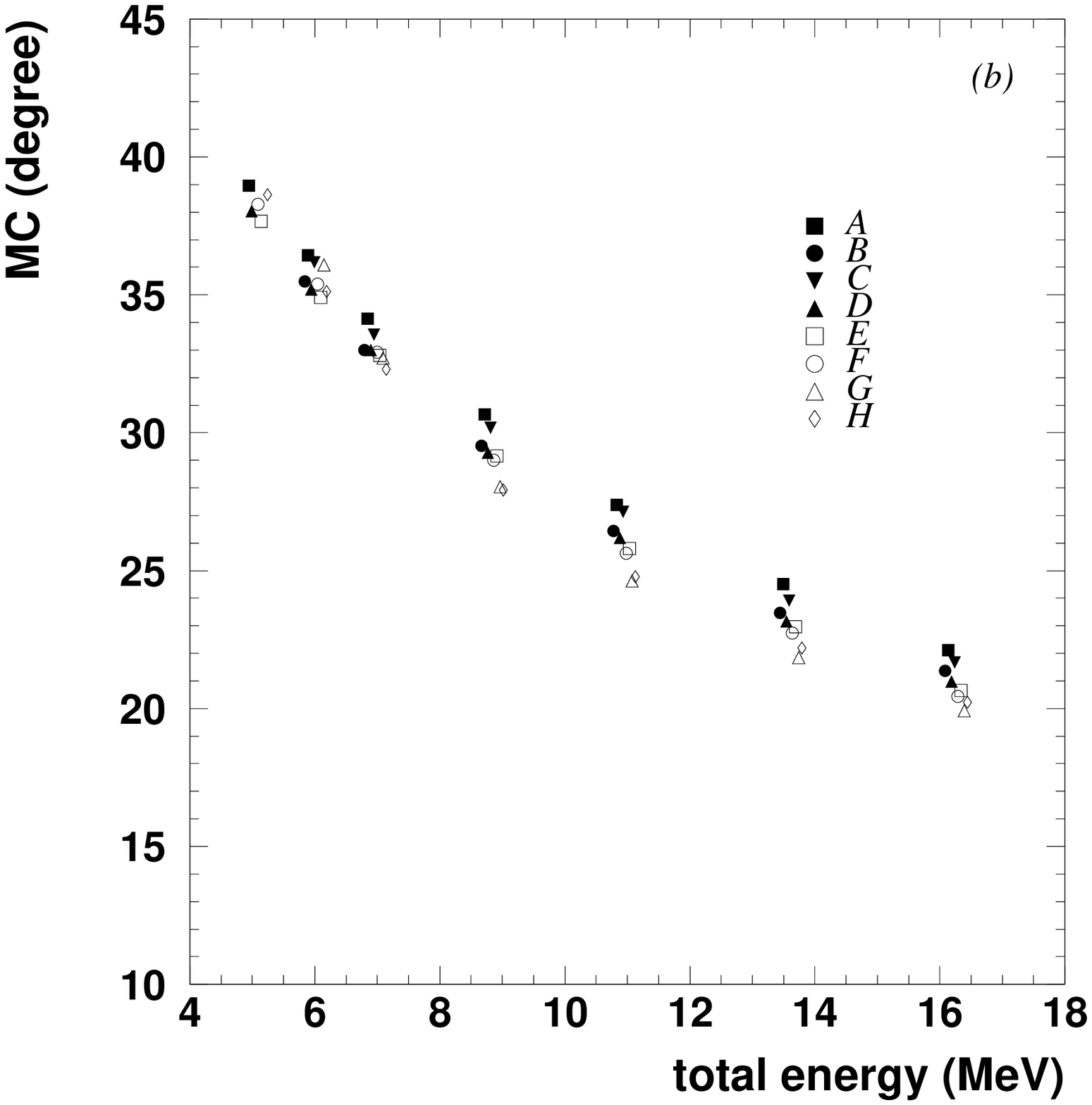}}
\resizebox{65mm}{!}{\includegraphics{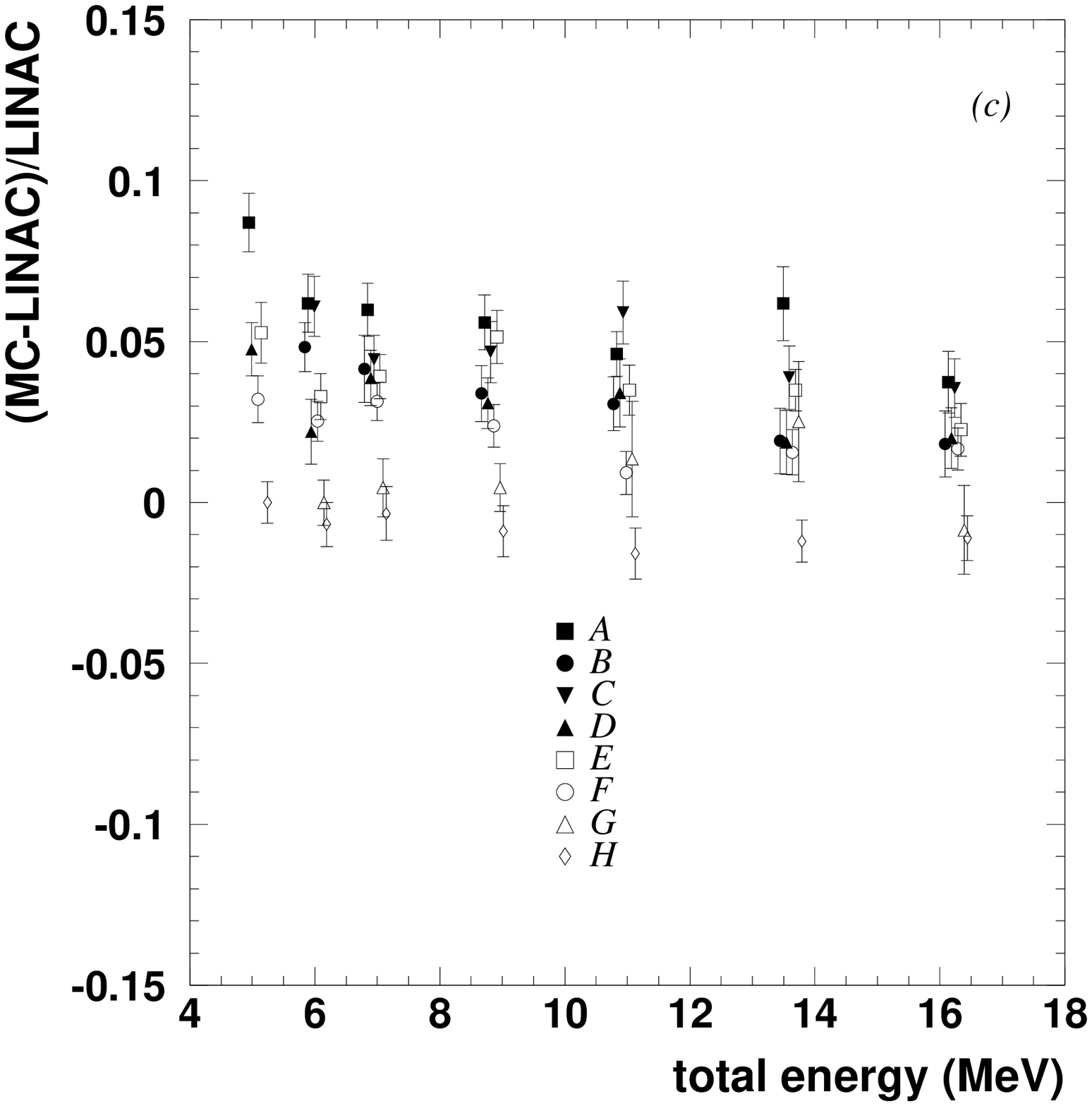}}
\resizebox{65mm}{!}{\includegraphics{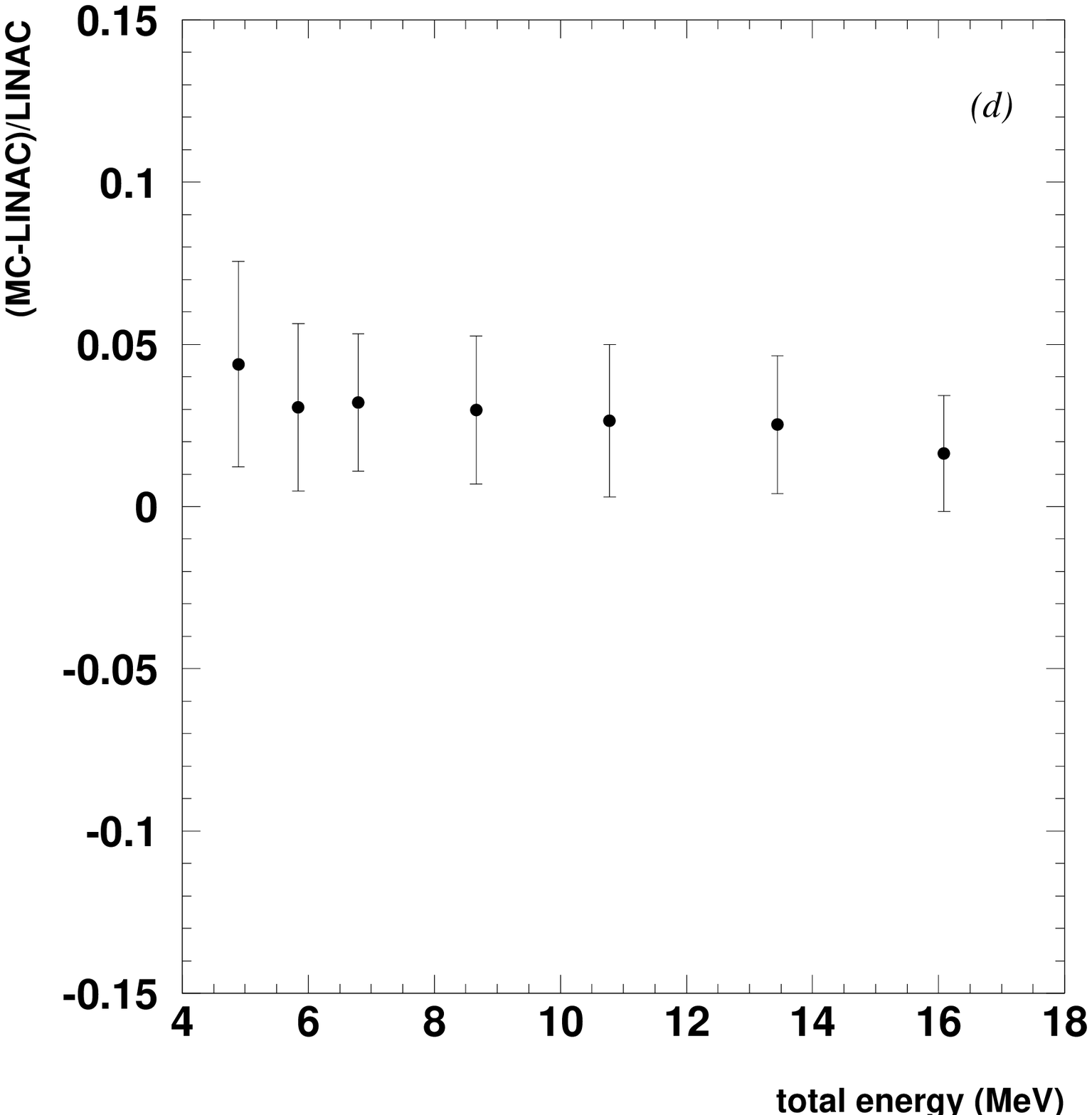}}
\end{center}
\caption{Energy dependence of angular resolution of (a) LINAC and (b) MC.
(c) is the difference between (a) and (b), (d) is averaged over positions. 
Errors as in Fig.~\protect{\ref{f:reVRES}}. 
A--H are defined in Tab.~\protect{\ref{t:dpos}}.} 
\label{f:reARES}
\end{figure}

Angular resolution is another quantity that is not tuned directly. 
Fig.~\ref{f:reANG} shows the opening angle between the reconstructed 
particle direction and the direction of beam injection for MC and LINAC
at a chosen position. 
In Fig.~\ref{f:reARES}(a) and (b), the angular resolution of LINAC and MC are 
displayed as functions of energy and position. 
Angular resolution is defined as the opening angle of a cone around the beam 
direction which contains 68\% of the reconstructed directions. 
Fig.~\ref{f:reARES}(c) shows the normalized offset between 
LINAC and MC resolution for all positions in the tank, 
Fig.~\ref{f:reARES}(d) the same quantity averaged over position. 
The relative difference between LINAC and MC becomes smaller for a slightly 
different choice of MC tuning parameters.  
The current choice of these parameters optimizes the uniformity of the energy 
scale throughout the detector volume. 

Tab.\ref{t:data} summarizes experimental resolutions of the Super-Kamiokande 
detector as observed on single electron LINAC events. 
Numbers for individual positions are extracted from 
experimental distributions like the ones shown in the figures 
\ref{f:reVTX}, \ref{f:reET}, and \ref{f:reANG} as described above. 
Entries in the table are averages over all LINAC positions. 
The spread observed between the various LINAC positions is reflected in 
the quoted errors (RMS).  

\begin{table}
\begin{tabular}{|c|c|c|c|}  
\hline
    total energy & energy resolution & angular resolution & vertex resolution \\
 (MeV) & (\%) & (degree)  & (cm) \\ 
\hline
4.89  & 20.9$\pm$0.6 & 36.7$\pm$0.2 & 182$\pm$21 \\
5.84  & 19.2$\pm$0.5 & 34.6$\pm$0.2 & 133$\pm$8 \\
6.79  & 18.0$\pm$0.3 & 32.0$\pm$0.1 & 108$\pm$5 \\
8.67  & 16.2$\pm$0.2 & 28.4$\pm$0.2 & 85$\pm$2 \\
10.78 & 14.7$\pm$0.3 & 25.3$\pm$0.2 & 73$\pm$2 \\
13.44 & 13.5$\pm$0.3 & 22.5$\pm$0.1 & 65$\pm$2 \\
16.09 & 12.6$\pm$0.3 & 20.6$\pm$0.1 & 50$\pm$2 \\
\hline
\end{tabular}
\caption{Experimental detector resolutions for Super-Kamiokande as derived 
from LINAC data}
\label{t:data}
\end{table}

\section{Conclusions}  \label{secCO}

The Super-Kamiokande detector is calibrated with an electron LINAC
for the energy range from 5 MeV to 16 MeV.
By this means the absolute energy scale for the solar neutrino analysis is 
known with an accuracy better than 1\%. 

MC simulation reproduces the energy resolution of the Super-Kamiokande 
detector to within 2\% and its angular resolution to better than 1.5~degrees 
for 10~MeV electrons. 
MC vertex resolution is well tuned to match the LINAC data. 

Various improvements of the current setup are in development and will soon be 
implemented. 
A permanent magnet for installation at the end of the beam pipe is 
currently being designed, 
which will allow bending of the LINAC electrons out of the -z direction. 

\section{Acknowledgement}  \label{secAK}

We gratefully acknowledge the cooperation of the Kamioka Mining and
Smelting Company.
This work was partly supported by the Japanese Ministry of Education,
Science and Culture and the U.S. Department of Energy.

\end{document}